\documentclass[12pt]{article}

\usepackage{fancyhdr,titlesec,geometry,ragged2e}
\usepackage[dvipsnames]{xcolor}
\usepackage[many]{tcolorbox}
\usepackage{layout}
\usepackage{lipsum}
\usepackage{hyperref}
\hypersetup{
  colorlinks,
  citecolor=red,
  linkcolor=blue,
  linktoc=all
}
\usepackage[T1]{fontenc}                            % Font Styling
\usepackage{lmodern,mathrsfs}

\usepackage[shortlabels]{enumitem}
\usepackage{mathtools,amssymb,amsfonts,amsthm,bm,mathrsfs}   % Math Presets
\usepackage{array,tabularx,booktabs}                % Table Presets
\usepackage{graphicx,wrapfig,float,caption,epsfig}         % Figure Presets
\usepackage{setspace,multicol}                      % Text Presets
\usepackage{tikz,tikz-cd,physics,cancel}            % Physics Presets

\usepackage{esint}
\allowdisplaybreaks

\usepackage{parskip}
\DeclareGraphicsExtensions{.pdf,.png,.jpg}
\usepackage{geometry}
\renewcommand{\thefootnote}{\arabic{footnote}}

\titleformat{\section}[hang]{\normalfont\large\bfseries}{\thesection.}{3mm}{}
\titlespacing{\section}{0mm}{15mm}{4mm}

\titleformat{\subsection}[hang]{\normalfont\bfseries}{\thesubsection}{2mm}{}
\titlespacing{\subsection}{0mm}{10mm}{4mm}

\titleformat{\subsubsection}[hang]{\normalfont\bfseries}{\thesubsubsection}{5mm}{}
\titlespacing{\subsubsection}{0mm}{5mm}{1mm}

\newtheoremstyle{thmstyle}
{5pt} % Space above
{3pt} % Space below
{\itshape} % Body font
{} % Indent amount
{\bfseries} % Theorem head font
{.} % Punctuation after theorem head
{0.5em} % Space after theorem head
{} % Theorem head spec (can be left empty, meaning `normal')

\theoremstyle{remark}{

}

\theoremstyle{thmstyle}{

\newtheorem{theorem}{Theorem}[section]
\newtheorem{proposition}[theorem]{Proposition}

\newtheorem{propcorollary}[theorem]{Corollary}
}

\numberwithin{equation}{section}
\renewcommand{\theequation}{\thesection.\arabic{equation}}
\newlength{\extraspace}
\newlength{\extraspaces}
\newcommand{\be}{\begin{equation}
\addtolength{\abovedisplayskip}{\extraspaces}
\addtolength{\belowdisplayskip}{\extraspaces}
\addtolength{\abovedisplayshortskip}{\extraspace}
\addtolength{\belowdisplayshortskip}{\extraspace}}
\newcommand{\ee}{\end{equation}}
\newcommand{\ba}{\begin{eqnarray}
\addtolength{\abovedisplayskip}{\extraspaces}
\addtolength{\belowdisplayskip}{\extraspaces}
\addtolength{\abovedisplayshortskip}{\extraspace}
\addtolength{\belowdisplayshortskip}{\extraspace}}
\newcommand{\ea}{\end{eqnarray}}
\newcommand{\bas}{\begin{eqnarray*}
\addtolength{\abovedisplayskip}{\extraspaces}
\addtolength{\belowdisplayskip}{\extraspaces}
\addtolength{\abovedisplayshortskip}{\extraspace}
\addtolength{\belowdisplayshortskip}{\extraspace}}
\newcommand{\eas}{\end{eqnarray*}}

\newcounter{subequation}[equation]
\makeatletter

\expandafter\let\expandafter
\reset@font\csname reset@font\endcsname

\def\subeqnarray{\arraycolsep1pt
    \def\@eqnnum\stepcounter##1{\stepcounter{subequation}%
        {\reset@font\rm(\theequation\alph{subequation})}}
\jot5mm     \eqnarray}

\def\subarray{\arraycolsep1pt
    \def\@eqnnum\stepcounter##1{\stepcounter{subequation}%
        {\reset@font\rm(\alph{subequation})}}
\jot5mm     \eqnarray}

\makeatother

\renewcommand{\dd}{{\partial}}

\renewcommand{\bra}{\langle}
\renewcommand{\ket}{\rangle}
\newcommand{\ra}{\rightarrow}

\newcommand{\sspace}{\makebox[1cm]{ }}

\newcommand{\nonum}{\nonumber \\[1.5mm]}
\newcommand{\is }{&\!\!=\!\!&} % for eqnarray
\DeclarePairedDelimiterX\set[1]\lbrace\rbrace{#1}

\newcommand{\inv}{^{-1}}

\newcommand{\1}{\mbox{1\hspace{-.8ex}1}}

\newcommand{\lb}{\lambda}

\newcommand{\om}{\omega}

\newcommand{\vp}{\varphi}

\newcommand{\eps}{\epsilon}

\renewcommand{\th}{\theta}

\newcommand{\cD}{{\cal D}}

\newcommand{\cL}{{\cal L}}

\newcommand{\cP}{{\cal P}}

\newcommand{\bbR}{{\mathbb{R}}}
\newcommand{\R}{\mathbb{R}}

\newcommand{\mfD}{{\mathfrak{D}}}

\newcommand{\wg}[0]{{\rm g}}

\def\fnum@figure{\textbf{\figurename\nobreakspace\thefigure}}
\def\fnum@table{\textbf{\tablename\nobreakspace\thetable}}

\newcommand{\vtest}{%
  \mathord{% ensure math mode and grouping
    \renewcommand{\arraystretch}{0}%
    \begin{array}[t]{@{}c@{}l@{}}
      \\[-11pt]
    \rightharpoonup\\[-1pt]
    \mfD
    \end{array} 
    \kern\scriptspace
  }%
}

\newcommand{\cvtest}{%
  \mathord{% ensure math mode and grouping
    \renewcommand{\arraystretch}{0}%
    \begin{array}[t]{@{}c@{}l@{}}
    \mfD\\[0.8pt]
    \rightharpoondown
    \end{array} 
    \kern\scriptspace
  }%
}
\newcommand{\vlp}[1]{%
  \mathord{% ensure math mode and grouping
    \renewcommand{\arraystretch}{0}%
    \begin{array}[t]{@{}c@{}l@{}}
    \\[-10.8pt]
    \rightharpoonup\\[-2.1pt]
    L^{#1} & 
    \end{array} 
    \kern\scriptspace
  }%
}

\newcommand{\cvlp}[1]{%
  \mathord{% ensure math mode and grouping
    \renewcommand{\arraystretch}{0}%
    \begin{array}[t]{@{}c@{}l@{}}
    L^{#1}\\[-0.1pt]
    \rightharpoondown
    \end{array} 
    \kern\scriptspace
  }
}

\newcommand{\dth}[0]{\Delta_\theta}

\begin{document}

\begin{titlepage}

%footnotesymbols others than numbers
\renewcommand{\thefootnote}{\arabic{footnote}}
\makebox[1cm]{}
\vspace{2mm}

% The title
\begin{center}
\hypersetup{linkcolor=black}
\mbox{{\large \bf Wick rotation in the lapse, admissible complex metrics, }} 
\\[4mm]
\mbox{{\large \bf and foliation changing diffeomorphisms}}
\vspace{2.3cm}

{\sc Rudrajit Banerjee}\footnote[1]{email:
\ttfamily\href{mailto:rudrajit.banerjee@oist.jp}{\textcolor{black}{rudrajit.banerjee@oist.jp}}} 
{\sc and} 
\renewcommand{\thefootnote}{\fnsymbol{footnote}}
{\sc Max Niedermaier}\footnote{Corresponding author} 
\renewcommand{\thefootnote}{\arabic{footnote}}%
\footnote[2]{email:
\ttfamily\href{mailto:mnie@pitt.edu}{\textcolor{black}{mnie@pitt.edu}}}
\\[8mm]
{\small\sl $^1$Okinawa Institute of Science and Technology Graduate University,}\\
{\small\sl 1919-1, Tancha, Onna, Kunigami District}\\ 
{\small \sl Okinawa 904-0495, Japan}
\\[5mm]
{\small\sl $^2$Department of Physics and Astronomy}\\
{\small\sl University of Pittsburgh, 100 Allen Hall}\\
{\small\sl Pittsburgh, PA 15260, USA}
\vspace{16mm}

{\bf Abstract} \\[4mm]
\begin{quote}
  A Wick rotation in the lapse (not in time) is introduced that
  interpolates between Riemannian and Lorentzian metrics on real
  manifolds admitting a codimension-one foliation. The definition
  refers to a fiducial foliation but covariance under foliation
  changing diffeomorphisms can be rendered explicit in a
  reformulation as a rank one perturbation. Applied to scalar field
  theories a Lorentzian signature action develops a positive
  imaginary part thereby identifying the underlying complex metric
  as ``admissible''. This admissibility is ensured in non-fiducial
  foliations in technically distinct ways also for the variation
  with respect to the metric and for the Hessian. The Hessian of the Wick
  rotated action is a complex combination of a generalized Laplacian
  and a d'Alembertian, which is shown to have spectrum
  contained in a wedge of the upper complex half plane.
  Specialized to near Minkowski space the induced propagator differs from
  the one with the Feynman $i\eps$ prescription and on
  Friedmann-Lema\^{i}tre backgrounds the difference to a Wick
  rotation in time is illustrated. 
\end{quote}
\end{center}

\vfill
\setcounter{footnote}{0}
\end{titlepage}
%%%%%%%%%%%%%%%%%%%%%%%%%%%%%%%%%%%%%%%   
% ----------------------------------------------------------------------
%		Table of contents page
% ----------------------------------------------------------------------
%\thispagestyle{empty}
%\makebox[1cm]{}
%
%\vspace{-23mm}
%\begin{samepage}                   
%{\hypersetup{linkcolor=black}
%\tableofcontents}
%\end{samepage}
%
%\nopagebreak

\newpage
%%%%%%%%%%%%%%%%%%%%%%%%%%%%%%%%%%%%%%%%%%%%%%%%%%%%%%%%%%%%%%%%

%%%%%%%%%%%%%%%%%%%
%	New section 
%%%%%%%%%%%%%%%%%%%

\setcounter{equation}{0}

%%%%%%%%%%%%%%%%%%%%%%%%%%%%%%%%%%%%%%%%%%%%%%%%%%%%%%%%%%%%%%%%%%%%%

\section{Introduction} 

%%%%%%%%%%%%%%%%%%%%%%%%%%%%%%%%%%%%%%%%%%%%%%%%%%%%%%%%%%%%%%%%%%%%%

For field theories on curved non-stationary backgrounds the notion of a Wick
rotation is problematic. The proposed approaches include:
rank one deformations \cite{Candelas,VisserWick1}, complex analytic
metrics \cite{Moretti1,AnalyticHad,WickWrochna}, and
Vielbein formulations \cite{LS,Samuel,KSWick,VisserWick2}.
They have different range of applicability, limitations, and 
occasionally overlap. For example, a Wick rotation in time
may be limited to purely electric metrics \cite{Wickelectric};
much of the Vielbein analysis is so far pointwise without change of chart.  
A recent survey \cite{Wicked} deems none of the existing proposals
fully satisfactory.

Here we explore a notion of a Wick rotation on $1\!+\!d$ dimensional
real smooth manifolds $M$ that admit a codimension-one foliation
$t \mapsto \Sigma_t$ into $d$ dimensional leaves which are level surfaces
$T = t$ of a scalar function $T$. Throughout, the atlas of charts of the
manifold $M$ is kept real and merely 
some of the metric components are complexified. The diffeomorphism group
changing charts likewise remains real (and we take it to consist of
smooth maps connected to the identity that are
orientation- and boundary preserving). In adapted coordinates
$y^{\mu} = (t, x^a)$ the Lorentzian and Euclidean metrics to be
related may then be parameterized according to 
\be
\label{i0} 
ds^2_{\eps_g} = g^{\eps_g}_{\mu\nu}(y)dy^{\mu} dy^{\nu} = \eps_g N^2 dt^2 + 
\wg_{ab} (dx^a + N^a dt)(dx^b + N^b dt)\,,
\ee
where $N$ is the lapse, $N^a$ the shift, and $\wg_{ab}$ the metric 
on $\Sigma_t$. We collect these fields into a triple
$(N, N^a, \wg_{ab})_{\eps_g}$, where the subscript indicates
the signature of the metric reconstructed from these data. 
The sign of the signature parameter $\eps_g = \pm 1$
cannot be flipped along a real path in $[-1,1]$ without encountering
degenerate metrics. Instead, we use in a fiducial foliation 
a phase rotation in the lapse:
\be
\label{Nwick} 
(N, N^a, \wg_{ab})_{\eps_g} \mapsto
(i \eps_g^{-1/2} e^{- i \th}N, N^a, \wg_{ab})_{\eps_g} \,,
\quad \th \in [0,\pi)\,,
\ee
where $\sqrt{\eps_g} = +1, i$ for $\eps_g = 1, -1$. Crucially, the time  
coordinate remains real; it is the lapse field $N(t,x)$ in the
reference foliation that is complexified. The conventions are
such that starting from either initial signature the line element
after (\ref{Nwick}) is $ds^2_{\th} = - e^{-2 i \th} N^2 dt^2 +
\wg_{ab} (dx^a + N^a dt)(dx^b + N^b dt)$. Thus Lorentzian and
Euclidean signature are recovered by the $\th \ra 0^+$ and
$\th \ra \pi/2$ limits, respectively, irrespective of the initial
signature. 

The metric (\ref{i0}) and the hence the notion of the Wick rotation
(\ref{Nwick}) is manifestly invariant under diffeomorphisms $t' = \chi^0(t)$,
${x'}^a = \chi^a(t,x)$ that preserve the fiducial foliation.
More general diffeomorphisms will however mix the component fields
$N,N^a,\wg_{ab}$ nontrivially. Based on explicit formulas for
this mixing the Wick rotated triples can consistently be transferred
to foliations other than the fiducial one. The resulting complex metric
$g^{\th}$ is then defined in a fully covariant way, most
concisely as a rank one perturbation of the metrics (\ref{i0}). 
With this in place the usual notions of tensorial covariance can
be established. On the linearized level a lapse-Wick rotated
version of the algebra of surface deformations arises.  

Another desirable feature of a Wick rotation is to result
in damping integrands starting from a formal Lorentzian signature
functional integral. This leads to the admissibility criterion for
complex metrics (and the action under consideration)
proposed in \cite{LS,KSWick}.
Here we limit ourselves to a minimally coupled selfinteracting scalar
field theory. The signs in (\ref{Nwick})  are chosen such the resulting
complex metric is admissible in the chosen reference foliation.
Based on the above notion of tensorial covariance this will
continue to hold in all other foliations. A subtlety arises for
the energy momentum tensor as defined in terms of the variational
derivative of the action with respect to the metric. This turns
out to invoke reference metrics of different signature and
positivity of the action's deformation has to be established
along different lines. The upshot is that although the
  lapse-Wick rotation depends on a choice of reference foliation, the
  well-posedness of the resulting functional integral does not. 

In many quantum field theoretical computations the Hessian
of the action under consideration is central. In particular,
this holds for the widely used Functional Renormalization Group
\cite{ASPercbook,ASRSbook} in which Euclidean signature is paramount
in order to apply heat kernel methodology. For minimally coupled
scalar field theories the complexified Hessian that arises from the
lapse-Wick-rotation reads $-i \Delta_{\th}$, where
\be
\label{hessian} 
\Delta_{\th} = - \sin \th\, \cD_+ - i \cos \th \, \cD_-, \quad
\th \in (0,\pi)\,,
\ee
interpolates between the generalized Laplacian
$\cD_+ = - \nabla_{+}^2 + V$ and ($-i$ times) the d'Alembertian
$\cD_- = - \nabla_{-}^2 + V$ (for a nonnegative bounded smooth potential $V$).
Note that in general $[\cD_+, \cD_-] \neq 0$. Hence, even if
the spectra of $\cD_{\pm}$ are assumed to be known, information on
$\Delta_{\th}$'s spectrum is not immediate. Along different lines we
show that the spectrum of $-i\Delta_{\th}$ is contained in a wedge
of the upper half plane $- (\pi + \tilde{\th}) \leq |{\rm Arg}\lb |
\leq \tilde{\th}$, with $\tilde{\th}:=\min\set{\theta,\pi-\theta}$.  
This reflects yet another aspect
of the admissibility of the underlying complex metrics $g^{\th}$.

The paper is organized as follows. After introducing the
lapse-Wick-rotation (\ref{Nwick}) we study its interplay
with foliation changing diffeomorphisms in Section \ref{sec2_1}.
The reformulation as a complex rank one perturbation of the real
metrics (\ref{i0}) is presented in Section \ref{sec2_2_0}.
In Section \ref{sec3_1} we introduce two notions of admissibility of a
complex metric and show that the metrics arising by
lapse-Wick-rotation satisfy both. Finally, the rationale for
the spectral properties of the complexified Hessian is described
in Section \ref{sec3_2}. Some background material on foliations and
the $1+d$ block decomposition of differentials is collected in
Appendix \ref{appA}. In Appendix B we discuss the specialization
to Minkowksi and Friedmann-Lema\^{i}tre backgrounds.

%%%%%%%%%%%%%%%%%%%
%	New section 
%%%%%%%%%%%%%%%%%%%
\newpage
\section{Phase rotated lapse and foliation changing diffeomorphisms}
\label{sec2}

As outlined, we consider $1+d$ dimensional real, smooth manifolds $M$
that admit a co-dimension-one foliation, $I \ni t \mapsto \Sigma_{t}$,
see Appendix \ref{appA}. In addition, $M$ is assumed to be equipped with
a metric of the form
\be
\label{ADM} 
ds^2_{\eps_g} = g^{\eps_g}_{\mu\nu}(y)dy^{\mu} dy^{\nu} = \eps_g N^2 dt^2 + 
\wg_{ab} (dx^a + N^a dt)(dx^b + N^b dt)\,,
\ee
for both values of $\eps_g = \pm1$. For both signatures, the leaves
$\Sigma_t$ of the foliation are the level sets of a smooth submersion
$T:M\to \bbR$ (referred to as a temporal function). When $\epsilon_g=-1$,
$dT$ is taken to be  everywhere timelike and the (spacelike) leaves are
assumed to be Cauchy surfaces; the resulting Lorentzian manifolds are
globally hyperbolic. We are not aware of a concise
established term for the analogous $\eps_g =+1$ (Riemannian) manifolds.
For short,
we shall refer to the metric components in (\ref{ADM}) as the ADM
(Arnowitt-Deser-Misner) fields. These comprise a positive lapse $N>0$,
the shift $N^a$, and the positive definite spatial metric $\wg_{ab}$.
We collect these fields into a triple $(N, N^a, \wg_{ab})_{\eps_g}$,
where the temporal function is tacit, and the subscript indicates the
signature of the line element (\ref{ADM}) reconstructed from it. 

For any foliation $I \ni t \mapsto \Sigma_{t}$ with associated ADM
triple $(N, N^a, \wg_{ab})_{\eps_g}$, our proposed notion of Wick rotation is 
\begin{equation}
\label{lwick0} 
\mathfrak{w}_{\theta} : \quad 
(N, N^a, \wg_{ab})_{\eps_g} \mapsto (i\eps_g^{-1/2}
e^{-i\th} N, N^a, \wg_{ab})_{\eps_g}, \quad \th \in [0, \pi)\,, 
\end{equation}
where $\sqrt{\eps_g} = +1, i$ for $\eps_g = 1, -1$. 
This is such that, starting from a fiducial foliation, one obtains a complexified line-element
\begin{equation}
\label{lwick1}
ds^2_{\eps_g} \mapsto ds^2_{\th} = - e^{-2 i \th} N^2 dt^2  +
\wg_{ab} (dx^a + N^a dt)(dx^b + N^b dt)\,.
\end{equation}
The case $\eps_g =-1$  gives $N \mapsto e^{ - i \th} N$ and
relates a Lorentzian signature ADM metric at $\th =0$ to a complexified one
that becomes Euclidean for $\th = \pi/2$. The case $\eps_g = +1$
gives $N \mapsto i e^{- i\th} N$ and relates the original
Euclidean ADM metric at $\th= \pi/2$ to a complexified one
that becomes Lorentzian for $\th =0$. The second half
$(\pi/2,\pi)$ of the $\th$ interval is carried along for later use.

We write ${\rm Diff}(M)$ for the group of real diffeomorphisms
$U \ni (t,x) \mapsto (\chi^0(t,x),\chi^a(t,x)) = (t',{x'}^a) \in U'$
(for open neighborhoods $U,U'$) that are smooth, connected to the
identity, as well as orientation preserving. An important subgroup ${\rm Diff}(\{\Sigma\})
\subset {\rm Diff}(M)$ are the foliation preserving diffeomorphisms
of the form $t'=\chi^0(t), {x'}^a = \chi^a(t,x)$. They preserve the
leaves $\Sigma_t$ of the foliation, potentially changing their
time labeling. The line elements (\ref{ADM}) and (\ref{lwick1}) are
manifestly invariant under foliation preserving diffeomorphisms.
In particular, the lapse Wick rotation (\ref{lwick0}) does not
depend on the choice of coordinates used to describe the given
fiducial foliation. A relevant question is, what happens if the
foliation is changed? To address this question we limit ourselves
to foliations equivalent to the original one, that is, foliations
that can be reached by an actively interpreted diffomorphism in
${\rm Diff}(M)$. An explicit formula for the action of
such foliation changing diffeomorphisms on the ADM data
$(N, N^a,\wg_{ab})_{\eps_g}$ will guide the analysis.

%%%%%%%%%%%%%%%%%%%%%%%%%%%%%%%%%%%%%%%%%%%%%%%%%%%%%%%%%%%%%%%

\subsection{Foliation changing diffeomorphisms}
\label{sec2_1} 

The Wick rotation (\ref{lwick0}), (\ref{lwick1}) inevitably
refers to a fiducial foliation. The 1-forms entering, i.e.~$N dt,
e^a := dx^a + N^a dt$, $a =1, \ldots, d$, comprise a frame on $M$
which we dub the foliation frame. It
is manifestly a coordinate independent notion and thus invariant
under passively interpreted diffeomorphisms, as long as the foliation
(i.e.~the underlying temporal function $T$) is held fixed.
Upon transition to a different temporal function $T'$ whose level surfaces
define a new (equivalent) foliation $t' \mapsto \Sigma_{t'}$ the
foliation frame transforms in a nontrivial way. Writing
$(t',{x'}^a) = (\chi^0(t,x), \chi^a(t,x))$ for the actively
interpreted diffeomorphisms, the transformation law comes out as
\begin{align}
\label{frametransf}
N' dt' & =
\frac{N}{D_{\eps_g}} \Big[ C dt + \frac{\dd t'}{\dd x^a} e^a \Big]\,,
\nonumber
\\[4mm]
  {e'}^a & = X_b^a \Big[ e^b - \eps_g \wg^{bc} \frac{\dd t'}{\dd x^c}
    \frac{N^2}{D_{\eps_g}^2} \Big(C dt + \dfrac{\dd t'}{\dd x^d} e^d \Big) 
\Big]\,,
\end{align}  
where
\ba
\label{diffeoshorth} 
D_{\eps_g} \is 
\sqrt{ C^2+\eps_g N^2\dfrac{\dd t'}{\dd x^c} \dfrac{\dd t'}{\dd x^d} \wg^{cd}}\,,
\quad 
C = \dfrac{\dd t'}{\dd t} - \dfrac{\dd t'}{\dd x^c} N^c\,, 
\nonum
X_b^a \is \frac{\dd {x'}^a}{\dd x^b}- \frac{1}{C} \frac{\dd t'}{\dd x^b}
\Big(\dfrac{\dd {x'}^a}{\dd t} - \dfrac{\dd {x'}^a}{\dd x^d} N^d \Big)\,.
\ea
We refer to Appendix \ref{appA} for the block decomposition of the differentials;
the combinations (\ref{diffeoshorth}) will occur frequently and
always refer to a generic underlying diffeomorphism that is suppressed in the notation.  
For the derivation of (\ref{frametransf}), Appendix A of \cite{SvsCtensor}
may be consulted. The mathematical equivalence between active and passive
diffeomorphism transformations requires that  
\begin{equation}
\label{ADMprime}
ds^2_{\eps_g} = \eps_g {N'}^2 {dt'}^2  +
\wg'_{ab} (d{x'}^a + {N'}^a dt')(d{x'}^b + {N'}^b dt')\,.
\end{equation}
This fixes the transformation law for $\wg'_{ab}$ and after 
stripping off the coordinate 1-forms from $N' dt'$ and
${e'}^a$ one one obtains the transformation
law for the ADM triples $(N, N^a, \wg_{ab})_{\eps_g}$ themselves
\cite{SvsCtensor} 
\be
\label{transf} 
{\rm transf}_{\eps_g} : (N, N^a, \wg_{ab})_{\eps_g} \;\mapsto \;
   (N', {N'}^a, \wg'_{ab})_{\eps_g} \,,
   \ee
where 
\begin{subequations}
\label{tripletransf}
\begin{eqnarray}\label{tripletransfa}
N' \is \frac{N}{D_{\eps_g}}
\\ \label{tripletransfb}
   {N'}^a \is - \frac{1}{D_{\eps_g}^2} \bigg(
   \Big( \dfrac{\dd {x'}^a}{\dd t} - 
\dfrac{\dd {x'}^a}{\dd x^d} N^d \Big) C
+\eps_g N^2 \dfrac{\dd {x'}^a}{\dd x^d} 
\dfrac{\dd t'}{\dd x^c} \wg^{cd} \bigg) 
\\ \label{tripletransfc}
{\wg'}_{ab} \is 
\Big( \dfrac{\dd x^c}{\dd {x'}^a} + \frac{\dd t}{\dd {x'}^a} 
N^c \Big)
\Big( \dfrac{\dd x^d}{\dd {x'}^b} + \frac{\dd t}{\dd {x'}^b} 
N^d \Big) \wg_{cd} 
+\eps_g N^2 \dfrac{\dd t}{\dd {x'}^a} \dfrac{\dd t}{\dd {x'}^b} 
\,.
\end{eqnarray}
\end{subequations}

{\bf Remarks.} 

(i) Upon linearization $t' = t - \xi^0(t,x) + O((\xi^0)^2)$,
${x'}^a = x^a - \xi^a(t, x) + O((\xi^a)^2)$, $N' = N + \delta_{\xi} N$, etc.,
the transformations (\ref{tripletransf}) read
\ba
\label{transflinear}
\delta_{\xi} N \is (\dd_t - N^a \dd_a)(\xi^0 N)
+ (\xi^a + \xi^0 N^a) \dd_a N\,,
\nonum
\delta_{\xi} N^a \is \dd_t( \xi^a + \xi^0 N^a) -
      [\cL_{\vec {N}}(\vec{\xi} + \xi^0 \vec{N})]^a + \eps_g
      N^2 \wg^{ab} \dd_b \xi^0\,,
\nonum
\delta_{\xi} \wg_{ab} \is \xi^0 (\dd_t - \cL_{\vec{N}})\wg_{ab} +
\cL_{\vec{\xi} + \xi^0 \vec{N}} \wg_{ab}\,.
  \ea
These generate the `group' of infinitesimal Lagrangian gauge
transformations of a generally covariant system,
c.f.~\cite{Pons}. Augmented
by $\delta_{\xi} \phi = \xi^{\mu} \dd_{\mu} \phi$, they comprise
in particular the gauge transformations of the scalar field action
(\ref{Saction}) below. Note that the $\eps_g$ dependence now only
enters in the $\delta_{\xi} N^a$ gauge transformation. By analogy
to (\ref{transf}) we shall write ${\rm lintransf}_{\eps_g}
(N, N^a, \wg_{ab})_{\eps_g} = (\delta_{\xi} N, \delta_{\xi} N^a,
\delta_{\xi} \wg_{ab})_{\eps_g}$, with the understanding that the
version of the matching signature is used. Conversely, one should
interpret (\ref{tripletransf}) as the finite gauge transformations
characterizing a generally covariant system with metrics in ADM form. 

(ii) To elucidate the `group' structure of (\ref{transflinear}) the
vector field $\xi^{\mu}$ is reparameterized according to \cite{Pons}
\be
\label{transflinear2}
\xi^0 = \frac{\eps^0}{N}\,, \quad \xi^a = \eps^a - \frac{\eps^0}{N} N^a\,,
\ee
and the field independent $\eps^0(t,x), \eps^a(t,x)$ are treated
as the descriptors of the infinitesimal gauge transformation.
Writing $\delta_{\xi(\eps^0,\vec{\eps})} N$, etc.~for the gauge variations
(\ref{transflinear}) expressed in terms of $(\eps^0,\eps^a)$ a
lengthy computation shows
\ba
\label{transflinear3}
&& \delta_{\xi(\eps_1^0,\vec{\eps}_1)}\delta_{\xi(\eps^0_2,\vec{\eps}_2)}
-\delta_{\xi(\eps_2^0,\vec{\eps}_2)}\delta_{\xi(\eps_1^0,\vec{\eps}_1)}
= -\delta_{\xi(\gamma^0,\vec{\gamma})}\,,
\nonum
&& \gamma^0 = \gamma^0(\eps_1^0, \vec{\eps}_1;\eps_2^0, \vec{\eps}_2) =
\eps_1^a \dd_a \eps_2^0 - \eps_2^a \dd_a \eps_1^0\,,
\nonum
&& \gamma^a = \gamma^a(\eps_1^0, \vec{\eps}_1;\eps_2^0, \vec{\eps}_2) =
\eps_1^b \dd_b \eps_2^a - \eps_2^b \dd_b \eps_1^a - \eps_g
\wg^{ab} (\eps_1^0 \dd_b \eps_2^0 - \eps_2^0 \dd_b \eps_1^0)\,,
\ea
when acting on (local functionals of) $N, N^a, \wg_{ab}$. The exchange relations (\ref{transflinear3})
are known as the ``algebra of surface deformations''. They are clearly
model independent and will (re-)occur in the Lagrangian formulation
of any generally covariant system.%
\footnote{In a Hamiltonian formulation with only the secondary
constraints kept the Hamiltonian gauge variations need to be augmented by
terms corresponding to an ``equations motion symmetry'' in order to
obtain a closed algebra isomorphic to (\ref{transflinear3}).} 
We display them here in order to discuss the effect of the 
lapse-Wick rotation on them later on.

(iii) On the right hand sides of \eqref{tripletransfa}, \eqref{tripletransfb}
the new adapted coordinates associated  to the  temporal function $T' = t'$
occur as functions of the original ones. In order to interpret the
last relation in the same way the inversion formulas (\ref{inv2})
ought to be inserted. For readability's sake we retain  the
given expression \eqref{tripletransfc} as a shorthand. 

(iv) The maps (\ref{tripletransf}) are invertible, and
the formulas for the inverse transformations can be obtained simply by
exchanging `primed' with `unprimed' quantities (fields and coordinate
functions). 

(v) In addition to being highly nonlinear the transformation laws
(\ref{frametransf}), (\ref{tripletransf}) also depend on the signature
parameter. As in \eqref{lwick0} this reflects the fact that we
take real, signature dependent metrics and the
associated ADM triples as a starting point. On triples
$(\sqrt{\eps_g} N, N^a, \wg_{ab})$ the foliation changing
diffeomorphisms act in an $\eps_g$ independent way (formally given
by the ${\rm transf}_+$ formulas).  

(vi) In the lapse transformation law a consistent square root needs to be
taken. This is possible since we restrict attention to separately time
and space orientation preserving diffeomorphisms. As far as the ADM metrics
are concerned one could work with triples $(N^2, N^a, \wg_{ab})_{\eps_g}$
where only the square of the lapse enters. Then ${\rm transf}_{\eps_g}$
would act as in (\ref{tripletransf}) just with $(N^2)'$ given
by the square of the right hand side of (\ref{tripletransfa}). 
\medskip

Using (\ref{frametransf}) one can deduce the transformation laws
of covariant tensor components defined with respect to the foliation
frame. For example, for a co-vector $V_{\mu} dy^{\mu} =
v N dt + v_a e^a = v' N' dt' + v_a' {e'}^a$ one finds%
\footnote{The relations (\ref{covectortransf}), (\ref{vectortransf})
  correct typos in the corresponding formulas  (A.53), (A.52) of
  \cite{SvsCtensor}.}
\ba
\label{covectortransf} 
v' \is \frac{1}{D_{\eps_g}}\Big(C v + \eps_g N \dfrac{\dd t'}{\dd x^c} \wg^{cd} v_d
\Big)\,,
\quad v'_a = \Big( \frac{\dd x^b}{\dd {x'}^a} + \frac{\dd t}{ \dd {x'}^a}
  N^b \Big) v_b + N \frac{\dd t}{\dd {x'}^a } v\,.
\ea
The frame dual to $(N dt, e^a)$ in the reference foliation consists of
the vector fields $(N^{-1} e_0,\dd_a)$. There are analogous transformation
formulas under a change of foliation, which can be found in
Appendix A of \cite{SvsCtensor}. We shall only need the induced 
transformation formulas for the components of a vector
$V^{\mu} \dd/\dd y^{\mu} =
\eps_g \check{v} N^{-1} e_0 + \check{v}^a \dd_a =
\eps_g \check{v}' {N'}^{-1} e'_0 + \mbox{$\check{v}'$}^a \dd'_a$,
which read
\ba
\label{vectortransf} 
\!\!\!\!\!\!\!\!\check{v}' \!\is \!
\frac{1}{D_{\eps_g}}\Big(C \check{v} + \eps_g N \dfrac{\dd t'}{\dd x^a}
\check{v}^a \Big) \,,\;\;
\check{v}'^a = X^a_b \bigg[ \check{v}^b - \wg^{bc} \frac{\dd t'}{\dd x^c}
\frac{N}{D_{\eps_g}^2} \Big(
C \check{v} + \eps_g N \dfrac{\dd t'}{\dd x^d } \check{v}^d \Big)\bigg].\,\,
\ea
\smallskip 

We now perform a Wick rotation (\ref{lwick0})
in the original foliation, resulting in the complex metric
(\ref{lwick1}). As in (\ref{lwick0}) we combine the
complexified ADM fields again into a triple
$(N_{\th}:= e^{- i \th}N, N^a, \wg_{ab})_{-}$, with the $-$ subscript \
indicating that the associated geometry arises through (\ref{lwick1}),
i.e.~$ds_{\th}^2 = - N_{\th}^2 dt^2 + \ldots$. Next, we subject the fields
$N_{\th} := e^{- i \th} N, N^a, \wg_{ab}$ to a foliation changing diffeomorphisms.
The fields referring to the resulting equivalent foliation 
$I \ni t' \mapsto \Sigma'_{t'}$ are denoted by a prime. On account of the
sign convention in (\ref{lwick1}) we use the ${\rm transf}_-$ transformations with its
domain extended to allow for a complex lapse. This gives 
$(N'_{\th}, {N'}_{\th}^a, {\wg'}^{\th}_{ab})_- =
{\rm transf}_{-}(N_{\th},N^a,\wg_{ab})_-$ with 
\begin{subequations}
\label{lwick2}
\begin{eqnarray}
	N_{\th}' \is \frac{N_{\th}}{\sqrt{ C^2
- N_{\th}^2\dfrac{\dd t'}{\dd x^c} \dfrac{\dd t'}{\dd x^d} \wg^{cd}}}\,, 
\\[2mm]
{N'_{\th}}^a \is - \dfrac{
\Big( \dfrac{\dd {x'}^a}{\dd t} - 
\dfrac{\dd {x'}^a}{\dd x^d} N^d \Big) C
- N_{\th}^2 \dfrac{\dd {x'}^a}{\dd x^d} 
\dfrac{\dd t'}{\dd x^c} \wg^{cd} 
}{C^2 - N_{\th}^2 \dfrac{\dd t'}{\dd x^c} 
\dfrac{\dd t'}{\dd x^d} \wg^{cd}}\,,
\\[2mm]
{\wg'}^{\th}_{ab} \is 
\Big( \dfrac{\dd x^c}{\dd {x'}^a} + \frac{\dd t}{\dd {x'}^a} N^c \Big)
\Big( \dfrac{\dd x^d}{\dd {x'}^b} + \frac{\dd t}{\dd {x'}^b} 
N^d \Big) \wg_{cd} 
- N_{\th}^2 \dfrac{\dd t}{\dd {x'}^a} \dfrac{\dd t}{\dd {x'}^b} 
\,.
\end{eqnarray}
\end{subequations}
The last relation should be interpreted in the same way as
\eqref{tripletransfc}.

The fact that also ${N'_{\th}}^a$, ${{\wg}'}^{\th}_{ab}$ are now complex in 
general highlights the sense in which the Wick rotation (\ref{lwick0})
is foliation dependent. However, specializing (\ref{lwick2}) to foliation 
preserving diffeomorphisms one sees that the $N_{\th}$ dependence 
in ${N'}^a$ and $\wg'_{ab}$ drops out, while $N'_{\th} = e^{- i \th} N' = 
(\dd t'/\dd t)^{-1} N_{\th} = (\dd t'/ \dd t)^{-1} e^{-i\th} N$ holds 
iff $N' = (\dd t'/\dd t)^{-1} N$.
Hence, the definition (\ref{lwick1}) only depends on the foliation 
and not on the coordinatization of the hypersurfaces or their 
time labels. 

The linearization of (\ref{lwick2}) leads to gauge variations
that can be obtained from the $\eps_g = -1$ version of (\ref{transflinear})
simply by the substitution $N \mapsto N_{\th} = e^{-i \th} N$. In the
reparameterization (\ref{transflinear2}) we insist on keeping
$\xi^0, \xi^a$ real and therefore phase rotate the descriptor
$\eps^0$ according to $\eps^0 \mapsto \eps^{\th} := e^{- i \th} \eps^0$.  
The computation leading to (\ref{transflinear3}) then carries over and
results in
\ba
\label{transflinear4}
&& \delta_{\xi(\eps_1^{\th},\vec{\eps}_1)}\delta_{\xi(\eps^{\th}_2,\vec{\eps}_2)}
-\delta_{\xi(\eps_2^{\th},\vec{\eps}_2)}\delta_{\xi(\eps_1^{\th},\vec{\eps}_1)}
= -\delta_{\xi(\gamma^{\th},\vec{\gamma}_{\th})}\,,
\nonum
&& \gamma^{\th} =
\gamma^0(\eps_1^{\th}, \vec{\eps}_1;\eps_2^{\th}, \vec{\eps}_2) =
\eps_1^a \dd_a \eps_2^{\th} - \eps_2^a \dd_a \eps_1^{\th}\,,
\nonum
&& \gamma_{\th}^a = \gamma^a(\eps_1^{\th}, \vec{\eps}_1;\eps_2^{\th}, \vec{\eps}_2) =
\eps_1^b \dd_b \eps_2^a - \eps_2^b \dd_b \eps_1^a +
\wg^{ab} (\eps_1^{\th} \dd_b \eps_2^{\th} - \eps_2^{\th} \dd_b \eps_1^{\th})\,,
\ea
when acting on (local functionals of) $N, N^a, \wg_{ab}$. This is the
lapse-Wick rotated
algebra of surface deformations. It interpolates between the Lorentzian
($\eps_g =-1$) and the Euclidean $(\eps_g =+1)$ versions of
(\ref{transflinear3}) (with the extra $-i$ in the zero components
attributed to the lapse redefinition, $N_{\pi/2} = -i N$).
The infinitesimal version has the
advantage that the gauge variations $\delta_{\xi(\eps^{\th},\vec{\eps})}$
refer to a single reference foliation due to the $N,N^a$-dependent
redefinition (\ref{transflinear2}).

The finite transformations
(\ref{lwick2}) extend the gauge symmetry to all orders in
$\xi^0, \xi^a$. By construction they form directly a group under
composition, but one needs to keep track of the three foliations invoked,
$\{\Sigma_t\} \stackrel{\chi_1}{\longrightarrow} \{\Sigma'_{t'}\}
\stackrel{\chi_2}{\longrightarrow} \{\Sigma''_{t''}\}$,
where $\chi_1 \circ \chi_2$ consistently maps $\{\Sigma_t\}$
to $\{\Sigma''_{t''})$. We summarize the key properties of (\ref{lwick2})
as follows.
\smallskip

\begin{proposition} \label{Wickmetric} The lapse Wick rotated metric
$g^{\th}_{\mu\nu} dy^{\mu} dy^{\nu}
  = - N_{\th}^2 dt^2 + \wg_{ab} (dx^a + N^a dt) (dx^b + N^b dt)$
  in the fiducial foliation $t \mapsto \Sigma_{t}$ gives
in a new (equivalent) foliation $t' \mapsto \Sigma'_{t'}$ 
rise to ${{g'}^{\th}}_{\!\!\!\!\mu\nu} d{y'}^{\mu} d{y'}^{\nu} =
- {N'_{\th}}^2 dt'^2 
+ {\wg'}^{\th}_{ab} (d{x'}^a + {N'_{\th}}^a dt') (d{x'}^b + {N'_{\th}}^b dt')$.  
This is such that
\begin{equation}
\label{lwick3}
g^{\th}_{\mu\nu} dy^{\mu} dy^{\nu} =
{{g'}^{\th}}_{\!\!\!\!\mu\nu} d{y'}^{\mu} d{y'}^{\nu} \,.
\end{equation}
We shall refer to (\ref{lwick3}) as the complexified metric
defined by lapse Wick rotation. It is invariantly defined
with respect to passive and active diffeomorphisms but depends
on the choice of fiducial foliation.  
\end{proposition} 

{\it Proof of Proposition \ref{Wickmetric}.} Viewing (\ref{lwick2})
as a definition only (\ref{lwick3}) needs to be shown. This can be
established by a lengthy direct computation.
\qed 

\bigskip
{\bf Wick rotation in non-fiducial foliations.} So far, the Wick rotation
(\ref{lwick0}) only acted in the arbitrarily chosen but then  fixed
fiducial foliation. The result was then transplanted to other
foliations by a foliation changing diffeomorphism. Formalizing
this construction, one can define a Wick rotation in a non-fiducial
foliation by the alternative expressions
\ba
\label{lwick4}
\mathfrak{w}'_{\th} &:= & {\rm transf}_- \circ \mathfrak{w}_{\th}
\circ ({\rm transf}_-)^{-1} \,,
\nonum
\mathfrak{w}'_{\th} &:= & {\rm transf}_+ \circ \mathfrak{w}_{\th}
\circ ({\rm transf}_+)^{-1} \,.
\ea
Here, $\mathfrak{w}'_{\th}$ acts on the real triples
$(N', {N'}^a, \wg'_{ab})_-$ and $(N', {N'}^a, \wg'_{ab})_+$,
respectively, of a matching signature metric in a non-fiducial
foliation. In the second ${\rm transf}_{\eps_g}$ map its
action is extended to allows for a complex lapse. 
In the notation (\ref{lwick2}) the result
is $\mathfrak{w}'_{\th} (N', {N'}^a, \wg'_{ab})_- =
(N'_{\th}, {N'}_{\th}^a, {\wg'}^{\th}_{ab})_-$ and
$\mathfrak{w}'_{\th} (N', {N'}^a, \wg'_{ab})_+ =
(i N'_{\th}, {N'}_{\th}^a, {\wg'}^{\th}_{ab})_+$. 
Since $(iN', {N'}^a, \wg'_{ab})_+ = (N', {N'}^a, \wg'_{ab})_-$
and $(i N'_{\th}, {N'}_{\th}^a, {\wg'}^{\th}_{ab})_+=
(N'_{\th}, {N'}_{\th}^a, {\wg'}^{\th}_{ab})_-$, both
variants of (\ref{lwick4}) are consistent; we keep both so
as to be able to work with real (signature dependent)
triples before Wick rotation. 

The cases $\th = \pi/2, 0$ are of particular interest and define
a {\it Wick flip}.  Specializing the defining relations in the
fiducial foliation $\mathfrak{w}_{\th}(N, N^a, \wg_{ab})_- =
(e^{- i \th} N, N^a, \wg_{ab})_-$, $\mathfrak{w}_{\th}(N, N^a, \wg_{ab})_+ =
(i e^{- i \th} N, N^a, \wg_{ab})_+$, to these cases one has 
\ba
\label{lwick5}
\mathfrak{w}_{\pi/2} (N, N^a, \wg_{ab})_- \is
( - i N, N^a, \wg_{ab})_- = (N, N^a, \wg_{ab})_+\,,
\nonum
\mathfrak{w}_{\pi/2} (N, N^a, \wg_{ab})_+ \is
(N, N^a, \wg_{ab})_+\,,
\nonum
\mathfrak{w}_0 (N, N^a, \wg_{ab})_- \is
(N, N^a, \wg_{ab})_- \,,
\nonum
\mathfrak{w}_0 (N, N^a, \wg_{ab})_+ \is
(i N, N^a, \wg_{ab})_+ = (N, N^a, \wg_{ab})_-\,.
\ea 
Note that $\mathfrak{w}_{\pi/2}^2 = \mathfrak{w}_{\pi/2}$,
$\mathfrak{w}_{0}^2 = \mathfrak{w}_{0}$, and
$\mathfrak{w}_0 \mathfrak{w}_{\pi/2}  = \mathfrak{w}_0$, 
$\mathfrak{w}_{\pi/2}\mathfrak{w}_0 = \mathfrak{w}_{\pi/2}$.
Clearly, the ${\rm transf}_+$ version of (\ref{lwick4}) is
trivial for $\mathfrak{w}_{\pi/2}$ while the ${\rm transf}_-$
version of (\ref{lwick4}) is trivial for $\mathfrak{w}_{0}$. 
The other two relations are 
\ba
\label{lwick6}
\mathfrak{w}'_{\pi/2} &:= & {\rm transf}_- \circ \mathfrak{w}_{\pi/2}
\circ ({\rm transf}_-)^{-1} \,,
\nonum
\mathfrak{w}'_0 &:= & {\rm transf}_+ \circ \mathfrak{w}_0
\circ ({\rm transf}_+)^{-1} \,,
\ea 
and extend the Wick flip to non-fiducial foliations. 
Explicitly, $\mathfrak{w}'_{\pi/2}(N',{N'}^a, \wg'_{ab})_- =
(N',{N'}^a, \wg'_{ab})_+$, and $\mathfrak{w}'_0(N',{N'}^a, \wg'_{ab})_+ =
(N',{N'}^a, \wg'_{ab})_-$.  
%%%%%%%%%%%%%%%%%%%%%%%%%%%%%%%%%%%%%%%%%%%%%%%%%%%%%%%%%%%%%%%%

\subsection{Complexified metric as a rank one perturbation}
\label{sec2_2_0}

In the fiducial
foliation the complexified metric can trivially be interpreted as
a rank one deformation of the original one. Writing, in adapted coordinates, $g^{(\eps_g)}_{\mu\nu} dy^{\mu} dy^{\nu}= \eps_g N^2 dt^2 +
\wg_{ab} e^a e^b$ and $g^{\th}_{\mu\nu} dy^{\mu} dy^{\nu} = - N_{\th}^2 dt^2
+ \wg_{ab} e^a e^b$ one has 
\begin{equation}
\label{deform1} 
g^{\th}_{\mu\nu}dy^\mu dy^\nu  = g^{(\eps_g)}_{\mu\nu}dy^\mu dy^\nu - (\eps_g + e^{-2 i \th})
N^2 dTdT\,,
\end{equation}
with $dT$  the differential of the temporal function of the foliation.
In other (primed) coordinates associated with another temporal function $T'$,
we seek to compare ${g'}^{\th}_{\!\mu\nu} {dy'}^{\mu} {dy'}^{\nu}$ as in
Proposition \ref{Wickmetric} with ${g'}^{(\eps_g)}_{\mu\nu}
{dy'}^{\mu} {dy'}^{\nu}$ from the right hand side of (\ref{ADMprime}).
One might guess that the deformation term in the new foliation arises
simply by placing `appropriate primes' on the original deformation,
i.e.~$N'^2 dT' dT'$. However, this is not the case, the correct assertion
being
\begin{proposition} \label{Wickrankone} 
The lapse Wick rotated metric, defined with respect to a fiducial foliation
in \eqref{lwick0}, is a rank one perturbation with a metric dependent covector
field. In any foliation equivalent to the fiducial one, 
\begin{equation}
\label{deform4} 
{g'}^{\th}_{\mu\nu} = {g'}^{(\eps_g)}_{\mu\nu} - (\eps_g + e^{-2 i \th})
\big( v' N' \dd'_{\mu}t' + v_a' {e'}^a_{\mu} \big)
\big( v' N' \dd'_{\nu}t' + v_a' {e'}^a_{\nu} \big)\,,
\end{equation} 
where with the notation from (\ref{diffeoshorth}) 
\begin{equation}
\label{deform3}
v' = 
\frac{C}{D_{\eps_g}}\,, \quad v_a' = N \frac{\dd t}{\dd {x'}^a} \,.
\end{equation}
Here, the $\dd t/\dd {x'}^a$ term should again be interpreted   
in terms of $t'$ via the inversion formula in (\ref{inv2}). 
\end{proposition}
{\it Proof.} The origin of the expressions for $(v', v_a')$ is simply as
the image of $v N dt + v_a e^a =
v' N' dt' + v_a' {e'}^a$ for $v=1, v_a =0$, using (\ref{covectortransf}).  
The last identity reaffirms the mathematical equivalence
between passive and active diffeomorphism transformations,
for the perturbing covector field. Since the latter is already known
to hold for the unperturbed metric via (\ref{ADMprime}) and
the Wick rotated one via Prop.~\ref{Wickmetric} it follows that
\begin{eqnarray} 
\label{deform2} 
&& - {N'_{\th}}^2 dt'^2 
+ {\wg'}^{\th}_{ab} (d{x'}^a + {N'_{\th}}^a dt') (d{x'}^b + {N'_{\th}}^b dt') 
\nonum
&& \quad = \eps_g {N'}^2 d{t'}^2 + \wg'_{ab} {e'}^a {e'}^b -
(\eps_g + e^{-2 i \th}) (v' N' dt' + v_a' {e'}^a)^2\,.
\end{eqnarray}   
Upon stripping off the coordinate differentials $d{y'}^{\mu}$
one obtains (\ref{deform4}). 
\qed

{\bf Remarks.} 

(i) The identity (\ref{deform2}) can also be verified by a lengthy direct
computation, using the formulae from Appendix A of \cite{SvsCtensor}.  
Note that the phase $e^{-i \th}$ enters the defining relations \eqref{lwick2}
highly nonlinearly on the left hand side while it appears only quadratically
on the right hand side. In particular, Eqs.~(\ref{deform4}), (\ref{deform2})
provide a satisfactory notion of general covariance for the lapse-Wick-rotated
metrics, i.e.~one not limited to defining ${g'}^{\th}_{\mu\nu}$ as the
image of $g^{\th}_{\mu\nu}$ under a generic diffeomorphism.

(ii) A notion of Wick rotation by a rank one deformation with a
complex coefficient $\lb$ has first been proposed in \cite{Candelas}. 
Their perturbing covector field $V_{\mu}$ is, however, taken as
a metric independent additional structure on the manifold.
For non-extreme values of $\lb$ the perturbed metric and
all concepts derived from it will depend on the choice of $V_{\mu}$. 
In the present setting the perturbing covector is itself
defined in terms of the metric data. Our complexified metric analogously
depends on the choice of fiducial foliation.

(iii) In \cite{Samuel,KSWick,VisserWick2} the complexification is
done in the internal metric of a Vielbein basis. That is, the 
Vielbein is kept real and merely the scalar diagonal 
coefficients are replaced by phases. In the present foliated setting
the natural Vielbein for (\ref{ADM}) is 
\ba
\label{vielbein1} 
E_I \is N^{-1} e_0 \,\eps_I + \eps_I^a \dd_a =
E_I^{\mu} \frac{\dd}{\dd y^{\mu}}\,,
\nonum
E^I \is \eps_g N dt \,\eps^I + \eps^I_a e^a = E^I_{\mu} dy^{\mu}\,,
\ea
where $E_I^{\mu} E_{\mu}^J = \delta_I^J$, $E_I^{\mu} E_{\nu}^I = 
\delta^{\mu}_{\nu}$, $I,J =0, \ldots, d$, and $g^{(\eps_g)}_{\mu\nu}
dy^{\mu} dy^{\nu} = \delta_{IJ}$ expresses the desired complete
diagonalization. The defining relations for the
component fields $(\eps_I, \eps_I^a)$ and $(\eps^I, \eps^I_a)$ can be
read off upon inserting (\ref{1d8}).  Applying the
lapse Wick rotation (\ref{lwick0}) to (\ref{vielbein1}) would preserve the
strict diagonalization at the expense of complexifying the Vielbeins. 
A better option is to retain the real Vielbeins (\ref{vielbein1})
and use the rank one formula (\ref{deform1}) to infer
\be
\label{vielbein2}
g^{\th}_{\mu\nu} E_I^{\mu} E_J^{\nu} = \delta_{IJ} -
(\eps_g + e^{-2 i \th}) \eps_I \eps_J\,.
\ee
This is no longer fully diagonal but has eigenvalues $(- e^{ - 2 i \th}, 1,
\ldots, 1)$. The transformation formulas (\ref{frametransf}) can be used
to deduce the induced behavior of the $\eps^I, \eps^I_a$ under foliation
changing diffeomorphisms, and similarly for $\eps_I, \eps_I^a$. This retains
the covariance in a sense analogous to the rank one perturbations
(\ref{deform4}).

(iv) For later use we also prepare the counterpart of the rank
one deformation formula (\ref{deform2}), (\ref{deform4}) for the inverse metric.
In the fiducial foliation one has
\be
\label{deform5} 
g_{\th}^{\mu\nu}(y)\frac{\dd}{\dd y^{\mu}} \frac{\dd}{\dd y^{\nu}} =
g_{\eps_g}^{\mu\nu}(y)\frac{\dd}{\dd y^{\mu}} \frac{\dd}{\dd y^{\nu}}
- ( \eps_g + e^{+2 i \th}) N^{-2} e_0^2\,.
\ee
The image in a generic foliation can be found in parallel to
(\ref{deform4}), (\ref{deform3}) using (\ref{vectortransf}).
for $\check{v} =1, \check{v}^a =0$. 
\be
\label{deform6}
{g'}_{\!\!\th}^{\mu\nu}(y')\frac{\dd}{\dd {y'}^{\mu}} \frac{\dd}{\dd {y'}^{\nu}} =
{g'}_{\!\!\eps_g}^{\mu\nu}(y')\frac{\dd}{\dd {y'}^{\mu}} \frac{\dd}{\dd {y'}^{\nu}}
- ( \eps_g + e^{+2 i \th}) \big( \eps_g \check{v}' {N'}^{-1} e'_0 +
\mbox{$\check{v}'$}^a \,\dd'_a \big)^2 \,,
\ee
where
\be
\label{deform7}
\check{v}' = \frac{C}{D_{\eps_g}}\,,
\quad \mbox{$\check{v}'$}^a = - \frac{N C}{D_{\eps_g}^2} 
X_b^a \wg^{bc} \frac{\dd t'}{\dd x^c}\,,
\ee
with $X_a^b$ from (\ref{diffeoshorth})

%%%%%%%%%%%%%%%%%%%%%%%%%%%%%%%%%%%%%%%%%%%%%%%%%%%%%%%%%%%%%%%%%%%%
\newpage 
\section{Admissible metrics for scalar field theories.} 
\label{sec2_2}

A reasonable ``admissibility criterion'' for a complex metric
$g^{\theta}_{\mu\nu} dy^{\mu} dy^{\nu}$ on a real manifold is that the
classically interpreted exponential of the action entering the functional
integral is damping. This reasoning is tacit in numerous discussions
of Wick rotations, recent explicit accounts are \cite{KSWick,VisserWick2,LS}.
Taking Lorentzian signature as basic and writing $S_{\th} =
S_-|_{g \mapsto g^{\th}}$ for the complexified action, $e^{ i S_{\th}}$ should
be damping. That is, ${\rm Im} S_{\th} >0$, for some range of $\th>0$,
if $S_{\th=0} = S_-$ is the Lorentzian signature action.
For short, we call a complex metric $g^{\theta}_{\mu\nu} dy^{\mu} dy^{\nu}$
on a real manifold {\bf admissible for $\mathbf{S}$} if this condition
is met for the action $S$ under consideration. In a small $\th$ expansion the linear
response, $S_{\th} = S_- +  (\delta S_-/\delta g_{\mu\nu}) (g^{\th} - g)_{\mu\nu}
+ O(\th^2)$, relates to the energy-momentum tensor
$T_-^{\mu\nu} = - (2/\sqrt{g}) \delta S_-/\delta g_{\mu\nu}$,
of the Lorentzian theory. The condition ${\rm Im} S_{\th} >0$ is then
to $O(\th)$ typically satisfied if the energy momentum
tensor satisfies the weak energy condition (WEC). For short, we 
call a complex metric {\bf WEC admissible for $\mathbf{S}$}
if $\Im S_{\th}>0$ holds
to $O(\th)$ on account of the WEC condition for $S$.
Note that a given complex metric could be admissible for one
action but not for another, it a theory dependent concept,
in contrast to the model independent considerations of Section 2.
For definiteness we focus below on the action of a minimally
coupled selfinteracting scalar field. We expect however that
the lapse-Wick rotated metrics remain admissible for any system
on foliated metric manifolds whose Euclidean action is bounded
from below.

On a foliated manifold both criteria are manifestly coordinate independent
(invariant under passive diffeomorphisms) as long as the fiducial
foliation is kept fixed. Below we limit ourselves to self-interacting
scalar fields on a foliated background and address the admissibility 
of our lapse Wick rotated complexified metric in foliations other
than the fiducial one in which the rotation is defined. Somewhat
surprisingly, the analysis is conceptually different for the exact
Wick rotation and the version linearized in $\th$.

%%%%%%%%%%%%%%%%%%%%%%%%%%%%%%%%%%%%%%%%%%%%%%%%%%%%%%%
\subsection{Linearised and nonlinear admissibility} 
\label{sec3_1}

We prepare the minimally coupled scalar field action for both signatures
\ba
\label{Saction}
S_{\eps_g}[\phi,g] \is \eps_g\! \int\! dy \sqrt{\eps_g g} 
\Big\{ \frac{1}{2} g_{\eps_g}^{\mu\nu} \dd_{\mu} \phi \dd_{\nu} \phi + U(\phi)
\Big\} 
\nonum
\is \int\! dt \!\int_{\Sigma} \! d^dx \sqrt{\wg} 
\Big\{ \frac{1}{2 N} e_0(\phi)^2 + \frac{\eps_g}{2} N \wg^{ab} \dd_a \phi 
\dd_b \phi + \eps_g N U(\phi)\Big\}\,. 
\ea   
In the second line we display the $1\!+\!d$ form of the action 
in some fiducial foliation with metric data $(N, N^a,\wg_{ab})_{\eps_g}$. 
Further,  
$U(\phi)$ is a metric independent potential which we assume 
to be non-negative. The bi-transversal component of the energy
momentum tensor $T^{\eps_g}_{\mu\nu}$ is defined by projection
with a real vector $m^{\mu}$ satisfying $dt_{\mu} m^{\mu} =1$,
$m^{\mu} m^{\nu} g^{\eps_g}_{\mu\nu} = \eps_g N^2$. 
This gives 
\ba
\label{SEM} 
T^{\eps_g}_{\mu\nu} \is \frac{2\eps_g}{\sqrt{g}}
\frac{\delta S_{\eps_g}}{\delta g^{\mu\nu}}
= \dd_{\mu} \phi \dd_{\nu} \phi - \frac{1}{2} g_{\mu\nu} 
g^{\rho \sigma} \dd_{\rho} \phi \dd_{\sigma} \phi - g_{\mu\nu} U(\phi)\,, 
\nonum
N^{-2} m^{\mu} m^{\nu} T^{\eps_g}_{\mu\nu} \is \frac{1}{2 N^2} e_0(\phi)^2 - 
\frac{\eps_g}{2} \wg^{ab} \dd_a \phi \dd_b \phi - \eps_g U(\phi)\,, 
\ea
where we momentarily omit the $\eps_g$ sub/superscripts on the metric
for readability's sake. One sees that $m^{\mu} m^{\nu} T^{-}_{\mu\nu} \geq 0$,
so Lorentzian signature scalar field theories with a non-negative potential
satisfy the WEC.

The action $S_{\eps_g}$ is manifestly invariant under foliation preserving
diffeomorphisms. In fact, each of the terms $N^{-2} e_0(\phi)^2$, 
$\wg^{ab} \dd_a \phi \dd_b \phi$, $U(\phi)$ is separately a 
scalar under ${\rm Diff}(\{\Sigma\})$ and the Wick rotation 
(\ref{lwick0}) can unambiguously be applied. Explicitly, 
we define in the fiducial foliation the lapse Wick rotated action
by
\ba 
\label{swick1}
S_{\th}[\phi,g] &:=& S_-[\phi,g]\big|_{N \mapsto e^{- i \th} N} 
= i S_+[\phi,g]\big|_{N \mapsto i e^{- i \th} N}
\nonum
\is
\cos \th S_-[\phi,g] + i \sin \th S_+[\phi,g] \,,
\ea
where $S_{\pm}$ are given by the second line in (\ref{Saction}). 
For $\th \in (0,\pi)$ one has $\Im[S_{\th}] > 0$ and the
generalized Boltzmann factor $e^{+i S_{\th}}$ in a functional
integral is damping. It is thus plain that the underlying complexified metric \eqref{lwick1} is admissible in the above sense in the
fiducial foliation. To linear order, $S_{\th} = S_- + i \th S_+ + O(\th^2)$.
Consistency with the WEC criterion requires that 
\begin{eqnarray} 
\label{swick2} 
S_+ &\stackrel{\displaystyle{!}}{=}&
\lim_{\th \ra 0^+} \frac{1}{i\th} \int\!\! dt d^dx \,
\frac{\delta S_-}{\delta g^{\mu\nu} }(g_{-}) \big(g^{\th} - g_{-}\big)^{\mu\nu}
\nonum
\is \lim_{\th \ra 0^+} \frac{ e^{2 i \th} -1}{2 i \th}
\int\!\! dt d^dx \, \sqrt{- g_{-}} N^{-2} \,T^{-}_{\mu\nu} m^{\mu} m^{\nu} \geq 0\,,
\end{eqnarray} 
where we used (\ref{deform5}) and the variational definition
of the energy momentum tensor. Inserting (\ref{SEM}) this is indeed an
identity. 

{\bf WEC admissibility in non-fiducial foliations.} The fiducial foliation can
of course be chosen arbitrarily and in this sense (\ref{swick2}) holds
in any foliation with its associated temporal function $T$.
One can, however, also ask if (\ref{swick2}) continues to hold
if the foliation is changed via the transformations
(\ref{tripletransf}). From the mathematical equivalence between active
and passive diffeomorphism transformations one expects
$T^{-}_{\mu\nu} m^{\mu} m^{\nu}$
not to be invariant (being the time-time component of a ${ 0 \choose 2}$ tensor)
and the issue is whether it remains positive. By comparing the second
lines of (\ref{Saction}) and (\ref{SEM}) one sees that both
$S_+$ and $T^{-}_{\mu\nu} m^{\mu} m^{\nu}$ contain the {\it sum} of the
temporal and the spatial gradient terms. By extension of
Proposition \ref{Wickmetric} these sums are scalars
under ${\rm transf}_+$ in (\ref{tripletransf}). However,
$T^{-}_{\mu\nu}$, stemming from 
the Lorentzian action should really be subjected to the ${\rm transf}_-$
transformations, and will then not be a scalar.  

It is instructive to compute explicitly the transformation law of the sum
and difference of the temporal and the spatial gradient parts
in the action $S_{\eps_g}$ based on the matching ${\rm transf}_{\eps_g}$
version of the transition formulas. Using the results from
Appendix A of \cite{SvsCtensor} one finds
\begin{subequations}
\begin{align}
\label{swick3a}  
& [{N'}^{-1} e_0'(\phi')]^2 + \eps_g {\wg'}^{ab} \dd'_a \phi' \dd'_b \phi' 
= 
[N^{-1} e_0(\phi)]^2 + \eps_g {\wg}^{ab} \dd_a \phi\dd_b \phi\,, 
\\[2mm]
\label{swick3b}
& [{N'}^{-1} e_0'(\phi')]^2 - \eps_g {\wg'}^{ab} \dd'_a \phi' \dd'_b \phi' 
= 
\dfrac{1}{ D_{\eps_g}^2 }
\bigg\{ \Big[ \frac{C}{N} e_0(\phi) + \eps_g N \frac{\dd t'}{\dd x^c} 
\wg^{cd} \dd_d \phi \Big]^2   
\nonum
& \sspace -  \eps_g \wg^{cd} 
\Big[ C \dd_c \phi - \frac{\dd t'}{\dd x^c} e_0(\phi) \Big] 
\Big[ C \dd_d \phi - \frac{\dd t'}{\dd x^d} e_0(\phi) \Big]
\bigg\}. 
\end{align}
\end{subequations}
The first combination occurs in the Lagrangian of $S_{\eps_g}$
and (\ref{swick3a}) confirms the expected scalar transformation law.
The sign flipped version occurs in the bi-transversal component of
the energy momentum  tensor (\ref{SEM}) and, as expected, does
not transform as a scalar under ${\rm transf}_{\eps_g}$. Relevant in
the present context is that the right hand side of (\ref{swick3b})
can be written so that for $\eps_g =-1$ is is manifestly
non-negative. Hence, when subjecting the second line of
(\ref{swick2}) to an active foliation changing diffeomorphism
of the inherited signature type, ${\rm transf}_-$, its
{\it value changes but it remains positive}. Hence,
WEC admissibility (for the scalar field action) is a
foliation-independent notion.  
\medskip

{\bf Admissibility in non-fiducial foliations.} The reason for
slightly belaboring the above point is that the situation is
conceptually different if the dependence on the phase $e^{ \pm i \th}$ is
treated exactly and no reference to the energy momentum tensor
of the original Lorentzian action is made. To frame the discussion
it is convenient to define $L(\phi,A) := \frac{1}{2} A^{\mu\nu} \dd_{\mu} \phi
\dd_{\nu} \phi + U(\phi)$, for any complex maximal rank matrix $A^{\mu\nu}$.
Then, in a given fiducial foliation $L(\phi,g_+)$ is the
Euclidean signature Lagrangian, $- L(\phi,g_-)$ is the Lorentzian
signature Lagrangian, and $- L(\phi,g_{\th})$ is the Lagrangian
of the complexified action (\ref{swick1}), excluding the complexified
measure term $\sqrt{ - g_{\th}}$. We interpret this measure term as
$\sqrt{ - g_{\th}} = e^{ - i \th} \sqrt{ \mp g_{\mp}} = e^{ -i \th}
N \sqrt{\wg}$. Taking the extra phase into account the Lagrangian
of the complexified action with the real $N \sqrt{\wg}$ measure is
$L_{\th} = - e^{ -i \th} L(\phi,g_{\th})$. In this notation the
relation (\ref{swick1}) reads
\be
\label{swick4}
- e^{ - i \th} L(\phi, g_{\th})(y) = - \cos \th \, L(\phi,g_- )(y) + i \sin \th\,
L(\phi, g_+)(y)\,,
\ee
where $y^{\mu} = (t, x^a)$ are local coordinates adapted to the
fiducial foliation. The interplay with non-fiducial foliations is described
by

\begin{proposition} \label{Ltransf}
  The Lagrangian $- e^{ - i \th} L(\phi, g_{\th})(y)$ of the
  complexified action is a scalar under the transformations (\ref{lwick2}),
$- e^{ - i \th} L(\phi, g_{\th})(y) = - e^{ - i \th} L(\phi', g_{\th}')(y')$.   
Explicitly, 
\begin{equation}
\label{swick5}
\frac{1}{2 N_{\th}^2} e_0(\phi)^2 - 
\frac{1}{2} \wg^{ab} \dd_a \phi \dd_b \phi - U(\phi)
=
\frac{1}{2 {N_{\th}'}^2} e'_0(\phi')^2 - 
\frac{1}{2} {\wg'}_{\th}^{ab} \dd'_a \phi' \dd'_b \phi' - U(\phi')\,,
\end{equation}
where $e_0' = \dd_t' - {N'}^a \dd_a'$ and ${\wg'}_{\th}^{ab}$ is the
inverse of ${\wg'}^{\th}_{ab}$ in (\ref{lwick2}c). Further,
\begin{equation}
\label{swick6}
- e^{ - i \th} L(\phi', g'_{\th})(y') = - \cos \th \, L(\phi',g'_- )(y')
+ i \sin \th\,L(\phi', g'_+)(y')\,.
\end{equation} 
In particular, the real and imaginary parts of
$- e^{ - i \th} L(\phi, g_{\th})$ are separately scalars
under the transformations (\ref{lwick2}). 
\end{proposition}

{\it Proof.} Since the inverse of the complexified metric
enters the `covariant' form of the action $S^{\th}[\phi,g] = S_-[\phi, g^{\th}]$ 
the assertion (\ref{swick5}) does not quite follow from (\ref{lwick3}). 
However, defining the inverses $g_{\th}^{\mu\nu}$ 
of $g^{\th}_{\mu\nu}$ and ${g'}_{\th}^{\mu\nu}$ of 
${g'}^{\th}_{\mu\nu}$ in the obvious way with respect to the
real vector field bases $\dd/\dd y^{\mu}$ and $\dd/\dd {y'}^{\mu}$,
respectively, it is clear that
\begin{equation}
\label{lwick12}
g_{\th}^{\mu\nu} \frac{\dd}{\dd y^{\mu}} \frac{\dd}{\dd y^{\mu}} 
= {g_{\th}'}^{\mu\nu} \frac{\dd}{\dd {y'}^{\mu}} \frac{\dd}{\dd {y'}^{\mu}} \,,
\end{equation}
will hold as well. This implies (\ref{swick5}).

The phase $e^{- i \th}$ occurs highly nonlinearly on the
right hand side of (\ref{swick5}). It is thus not immediate that
the latter can be decomposed as claimed on the right hand side of (\ref{swick6}). 
To see that this is the case, we return to (\ref{deform6}) and insert
it into the left hand side of (\ref{swick6}). In a first step this gives
\ba
\label{swick7} 
&& - e^{ - i \th} L(\phi', g'_{\th})(y') =
e^{ - i \th} \Big\{\! - \frac{1}{2} {g'_{\eps_g}}^{\!\!\mu\nu} \dd'_{\mu} \phi'
\dd'_{\nu} \phi' - U(\phi') 
\nonum
&&
\quad - \frac{1}{2}(\eps_g + e^{+2 i \th})
\big( \eps_g \check{v}' {N'}^{-1} e'_0(\phi') +
\mbox{$\check{v}'$}^a \,\dd'_a \phi'\big)^2 \Big\}\,.
\ea
By construction, either sign $\eps_g = \pm 1$ can be chosen to evaluate the
right hand side. Choosing $\eps_g = +1$ one finds
\ba
\label{swick8} 
&& - e^{ - i \th} L(\phi', g'_{\th})(y') =
i \sin \th \Big\{ \frac{1}{2} {g'_+}^{\!\!\mu\nu} \dd'_{\mu} \phi'
\dd'_{\nu} \phi' + U(\phi') \Big\} 
\nonum
&&
\quad - \cos \th \, \Big\{ \frac{1}{2} {g'_+}^{\!\!\mu\nu}
\dd'_{\mu} \phi' \dd'_{\nu} \phi'
- \big( \eps_g \check{v}' {N'}^{-1} e'_0(\phi') +
\mbox{$\check{v}'$}^a \,\dd'_a \phi'\big)^2 - U(\phi') \Big\}\,.
\ea
The first two terms in the second curly bracket can be simplified
using the $\th =0$, $\eps_g =+1$ version of (\ref{deform6})
in reverse. This yields (\ref{swick6}). 
\qed

In summary, also the nonlinear admissibility (for the scalar
field action) is a foliation-independent feature. 
In the context of our previous discussion of the WEC admissibility,
the result (\ref{swick6}) is somewhat surprising. While in
(\ref{swick2}) the imaginary part of the $O(\th)$ perturbation is
{\it not} a scalar under the inherited ${\rm transf}_-$ transformation,
the real and the imaginary parts in (\ref{swick6}) suddenly are. 
This is because the complex transformations (\ref{lwick2}) automatically
apply the matching transformations ${\rm trans}_{\pm}$ to the
definite signature parts of the quantities occurring on the
right hand side of (\ref{deform6}). As a consequence, after re-expressing
$- e^{ - i \th} L(\phi', g'_{\th})(y')$ in terms of the definite
signature $L(\phi',g'_-)$ and $L(\phi',g'_+)$ the latter
coincide with the images of $L(\phi, g_-)$ and $L(\phi, g_+)$
under the matching ${\rm transf}_-$ and ${\rm transf}_+$ transformations,
respectively. There is no inherited transformation law that is kept fixed
and results in a non-scalar transformation law. 

%%%%%%%%%%%%%%%%%%%%%%%%%%%%%%%%%%%%%%%%%%%%%%%%%%%%%%
\subsection{The complexified Hessian}
\label{sec3_2}

Next, we consider the Hessian defined by the quadratic part of the
action $S_{\th}$. The appropriate background-fluctuation split
is $\phi = \vp + f$, for a background $\vp$ and some $f \in C_c^{\infty}(M)$. 
We do not require $\vp$ to be on-shell for the reasons explained below. 
While on-shell backgrounds are commonly used for simplicity, they are
not mandatory in the background field formalism of functional integrals.
In particular, the Legendre effective action $\Gamma[ \bra f\ket, \vp]$
can consistently be defined for off-shell backgrounds. 

Expanding the action (\ref{swick1}) to quadratic order in $f$ one has  
\begin{equation}
\label{chess1} 
S_{\th}[\vp + f, g] = S_{\th}[\vp , g] - \int \!\!dt \!
\int_{\Sigma} \! d^dx \,N\sqrt{\wg}\, f \,i \Delta_{\th} \vp - \frac{1}{2}
\int \!\!dt \!\int_{\Sigma} \! d^dx \,N\sqrt{\wg} f \,i \Delta_{\th} f +
O(f^3)\,.
\end{equation}
The Hessian $- i \Delta_{\th}$ can be written in several
alternatively useful ways
\begin{align}
  \label{chess2}
  - i \Delta_{\th} & = - e^{-i \th}\big[
    - \nabla_{-}^2\big|_{N \mapsto e^{- i \th} N} + V \big]
  = - e^{ i \th} \nabla_t^2 + e^{ - i \th} \nabla_s^2 - e^{- i \th} V
\nonum
& = - \cos \th \, \cD_- + i \sin \th \, \cD_+\,,
\end{align} 
where $\cD_{\pm} := - \nabla_{\pm}^2 + V$, $V=U''(\varphi)$, are the
Euclidean/Lorentzian signature Hessians, respectively.
The first equality in (\ref{chess2}) from the first expression
  for $S_{\th}$ in (\ref{swick1}), with the extra phase stemming from the
  (originally positive) lapse term in the measure. For the second
  identity we decompose the familiar expression for the scalar Laplacian
into a temporal and a spatial part. Explicitly,  
\begin{align}
  \label{chess3}
  \nabla_{\!\eps_g}^2 & = (\epsilon_g g_{\eps_g})^{-1/2}
  \dd_{\mu} \big( (\epsilon_g g_{\eps_g})^{1/2}
g_{\eps_g}^{\mu\nu} \dd_{\nu} \big) 
\nonum
& = \eps_g \wg^{-1/2} N^{-1} e_0\big( \wg^{1/2} N^{-1} e_0 \big) 
+ \wg^{-1/2} N^{-1} \dd_a\big( N \wg^{1/2} \wg^{ab} \dd_b \big) 
=: \eps_g \nabla^2_t + \nabla^2_s\,. 
\end{align}
Here $e_0 = \dd_t - \cL_{\vec{N}}$ is the Lie derivative transversal to 
the leaves of the foliation. Note that the rightmost $e_0$ acts on
spatial scalars as $e_0(f) = \dd_t f - N^a \dd_a f$, while the next $e_0$ 
acts on a  $+1$ spatial density according to $e_0(\sqrt{\wg} f) =
\dd_t( \sqrt{\wg} f) - \dd_a( N^a \sqrt{\wg} f)$. In $1+d$ form the
diffeomorphism group 
acts nonlinearly according to the transformation formulas in
(\ref{tripletransf}) but for fixed signature parameter $\eps_g$,
-- the same in (\ref{tripletransf}) and (\ref{chess3})--, $\nabla^2_{\!\eps_g}$
will continue to map scalars to scalars. The temporal and spatial
parts individually are of course only invariant under foliation
preserving diffeomorphisms. The third version of $- i \Delta_{\th}$
in (\ref{chess2}) follows from the second by separating the real
and imaginary parts and using (\ref{chess3}) in reverse.   

The structure (\ref{chess2}) carries over to non-fiducial foliations 
on account of Prop.~\ref{Ltransf}.  

\begin{propcorollary} The complexified Hessian (\ref{chess2}) is invariant
  under the complex transformations (\ref{lwick2}), i.e.
  $\Delta_{\th}' = \Delta_{\th}$, in the respective local coordinates.
  Also in generic non-fiducial foliations it decomposes according to
  $ - i \Delta'_{\th}= - \cos \th \, \cD'_- + i \sin \th \, \cD'_+$,
  where $\cD'_{\pm}$ refer to $(N', {N'}^a, \wg'_{ab})_{\pm}$ and
are separately invariant, $\cD'_+ = \cD_+$, $\cD'_- = \cD_-$,   
with respect to ${\rm transf}_+$, ${\rm transf}_-$ in
(\ref{tripletransf}). 
\end{propcorollary} 

{\bf Remarks.}

(i) We do not require $\vp$ to be on-shell, i.e.~to be a solution of
$\Delta_{\th} \vp =0$. Imposing $\Delta_{\th} \vp =0$ for any fixed
$\th$ is unproblematic; its extension to all $\th$ requires however
$\cD_+ \vp = 0 = \cD_- \vp$ and thus would allow only simple
(e.g.~static) backgrounds. Instead, we leave $\vp$ generic and 
treat the potential $V = U''(\vp)$ that arises as a  given scalar
function on $M$.

(ii) Until now, we regarded all differential operators as tacitly acting
on $C_c^{\infty}(M)$, the smooth functions with compact support. 
In particular, both $\Delta_{\th}$ and $\Delta_{\pi - \th}$
can act on  $C_c^{\infty}(M)$. However, they are not adjoints
of each other on this domain. This can be fixed by enlarging the
domain to a subset $D(\Delta_{\th})$ of a Sobolev space. We omit
the detailed definitions \cite{Wickheatk} but note that as sets one has the
dense inclusions $C_c^\infty(M) \subseteq D(\dth)\subseteq L^2(M)$. 

(iii) In general $[\cD_+, \cD_-] \neq 0$. Hence, even if
the spectra of $\cD_{\pm}$ are assumed to be known, information on
$\Delta_{\th}$'s spectrum is not immediate.

The relevant result is \cite{Wickheatk}
%
%\footnote{The domain specification and the proof can also be found
%in the temporary arXiv entry, {\tt arXiv:2406.06047}.}

\begin{proposition} \label{sector}
Let $\Delta_{\th} = - \sin \th\, \cD_+
  - i \cos \th \, \cD_-$, $\th \in (0,\pi)$ be defined on the
 domain $D(\Delta_{\th})$ from the above remark (ii) for a
  nonnegative bounded smooth
  potential $V$. Then  
\begin{itemize}
\item[(a)]  The adjoint is given by $\Delta_{\th}^* = \Delta_{\pi - \th}$,
  including domains $D(\Delta_{\th}^*) = D(\Delta_{\pi - \th})$.  
\item[(b)] The spectrum of $\Delta_{\th}$ is contained in a wedge
of the left half plane, $|{\rm Arg} \lb| \geq \pi/2 + \tilde{\th}$,
with $\tilde{\th}:=\min\set{\theta,\pi-\theta}$.  
\end{itemize}   
\end{proposition}   

For the Hessian $-i \Delta_{\th}$ this means its spectrum lies in
a wedge of the upper half plane, $- (\pi + \tilde{\th}) \leq
{\rm Arg}(-i \lb) \leq \tilde{\th}$. Writing
$\frac{1}{2} f \cdot S_{\th}^{(2)}(\vp) \cdot f$ for the quadratic
part in (\ref{chess1}) this means that in a spectral representation
its imaginary part would be positive. The property (b) thus
codes yet another aspect of the admissibility of the underlying
complex metrics $g^{\th}$.

The property shown in Prop.~\ref{sector} is known as ``sectoriality'',
and allows the application of holomorphic operator calculus for the
(no longer self-adjoint or even symmetric) $\Delta_{\th}$. This can be
used to give rigorous meaning to desired objects like $e^{s \Delta_{\th}}$ and 
$(z - \Delta_{\th})^{-1}$, and their associated integral kernels, or regularized tracelog's in parallel to
the Euclidean case. Further, the strict Lorentzian limit is governed by the 
fact that $\lim_{\th \ra 0^+} {\rm Tr}[A\,
  e^{ s \Delta_{\th}}]$ is well-defined for any trace-class operator $A$
\cite{Wickheatk}.

%%%%%%%%%%%%%%%%%%%%%%%%%%%%%%%%%%%%%%%%%%%%%%%%%%%%%%%%%%%%%%%%%%%%%%%%%%
\section{Conclusions} 

A Wick rotation in the lapse, rather than in time, has been introduced
that interpolates between Lorentzian and Riemannian metrics of ADM form.
In contrast to other notions of Wick rotation the manifold (i.e.~its
coordinate atlases) stay real throughout. The lapse $N$ and hence the
ensued notion of Wick rotation depends on a choice of
fiducial foliation. Based on explicit formulas for the 
mixing of ADM triples $N,N^a,\wg_{ab}$ under foliation changing
diffeomorphisms, the initial Wick rotated triple can be transferred to
any other foliation. In the reformulation as a rank one perturbation 
a satisfactory notion of general covariance arises for the complexified
metrics. In particular, on a linearized level a lapse-Wick rotated
version of the algebra of surface deformations arises.

The resulting complex metrics are also ``admissible'' \cite{KSWick,LS}
in the sense of giving rise to damping integrands in an initially formal
Lorentzian signature functional integral. This of course depends
on the action under consideration and is demonstrated in detail for the   
action of a minimally coupled selfinteracting scalar field. We expect
it to carry over to any system (on a foliated metric manifold) whose
Euclidean action is bounded from below.
This admissibility
has several aspects: (i) for the energy-momentum tensor,
i.e.~the linear response under a variation of the metric. (ii) for
the complexified action itself. (iii) for the spectrum of its Hessian,
i.e.~the operator governing the part quadratic in fluctuations
of the matter field. For scalar field theories all three notions
of `admissibility' were seen to be satisfied. When specialized
  to Minkowski space one finds that the lapse-Wick rotation
does in the limit $\theta\to 0^+$ not induce the usual
    $i\epsilon$-prescription for the Feynman propagator, but
(as detailed in Appendix B)
  an improved variant
  introduced by Zimmermann \cite{Zimmermann}. The admissibility then
  manifests itself in the absolute (rather than conditional) convergence
of the relevant Feynman integrals. 

For definiteness we considered here only the scalar Hessian. The
lapse-Wick rotation carries over to actions with
vectorial, tensorial, or ghost degrees of freedom and the
associated Hessians. These Hessians can normally be decomposed into
generalized Lichnerowicz Laplacians/d'Alembertians.  
Clarifying the spectral properties of their lapse-Wick-rotated
versions would pave the way for a construction of the associated analytic
semigroups along the lines of \cite{Wickheatk}. For Euclidean signature
heat semigroups associated with Lichnerowicz Laplacians are widely
used to investigate
the quantum theory of gauge fields and gravity, often in combination
with the non-perturbative Functional Renormalization Group
\cite{ASPercbook,ASRSbook}. We see no principle obstruction
to such a generalization, which would allow one to explore the near
Lorentzian regime of such computations in an  apples-to-apples
comparison.

Finally, we mention the construction of a lapse-Wick-rotated Synge
function (one-half of the geodesic distance-squared between nearby points)
as a desideratum. A straightforward adaptation of the known constructions
\cite{Synge1,Synge2} would require locally analytic manifolds. This is at
odds with the real manifold  setting adopted here and presumably
also not necessary for the existence of a lapse-Wick-rotated Synge
function. An asymptotic expansion for it is known \cite{Wickheatk} but 
for use in off-diagonal expansions of the semigroups kernels
exact solutions are needed. Control over the
lapse-Wick rotated off-diagonal kernels would also allow for
the application of heat kernel techniques beyond (selfconsistently
improved) one loop level on curved backgrounds \cite{Gersd}
and their Lorentzian limits.  
\vspace{5mm} 

{\bfseries Acknowledgments.} We would like to thank
R.~Percacci and  F.~Saueressig for fruitful discussions related to
this topic and the Radboud University Nijmegen for
hospitality on several occasions. R.B also acknowledges support of the
Institute Henri Poincar\'{e} (UAR 839 CNRS-Sorbonne Universit\'{e}),
and LabEx CARMIN (ANR-10-LABX-59-01) during the `Quantum and classical
fields interacting with geometry' thematic program.

% ----------------------------------------------------------------------
%		Appendices
% ----------------------------------------------------------------------
\newpage
\appendix

%%%%%%%%%%%%%%%%%%%%%%%%%%%%%%%%%%%%%%%%%%%%%%%%%%%%%%%%%%%%%

\section{Foliation geometry} 
\label{appA}
\setcounter{equation}{0}

In this appendix we set our notation and collect a few basic notions
of foliation geometry in relation to foliation changing diffeomorphisms,
as needed in the main text. Throughout $M$ is a $1\!+\!d$ dimensional
topological manifold (locally Euclidean and Hausdorff) that is:
smooth, connected, orientable, $2^{\rm nd}$ countable,
and without boundary. We allow it to be noncompact.

{\bf Equivalent foliations.}
No metric structure is assumed in this part. A {\itshape co-dimension-one
  foliation} of $M$ is a collection $\set{\Sigma_\alpha}_{\alpha\in A}$ of
connected disjoint subsets of $M$ such that:
(i) $M=\cup_{\alpha \in A}\Sigma_\alpha$, and
(ii)every point in $M$ has a neighborhood $U$ and a system of local coordinates 
$y = (y^0,y^1,\ldots, y^d) : U \ra \R^{1+d}$, such that for each leaf
$\Sigma_\alpha$, if $\Sigma_\alpha\cap U \neq \emptyset$, then its
local coordinate image is a $y^0=\text{const.}$ slice of the chart range.
Such a (non-unique) coordinate system is said to be {\itshape adapted
  to the foliation}. Criteria for a manifold to admit such a structure
can be found in \cite{ThurstonFoli} and the references therein. Here we
assume that $M$ admits a co-dimension-one foliation given by the level
sets of a smooth submersion $T:M\to \bbR$ (in particular
$dT\neq 0$ everywhere).%
\footnote{In metric geometry $T$ corresponds to a temporal 
function and the associated foliations are vorticity-free,
see below. 
}
The foliation can then be parameterized as $\set{\Sigma_t}_{t\in I}$,
$I\subseteq \bbR$ is the range of $T$, and $\Sigma_t:=T\inv(\set{t})$;
by slight abuse of notation we
often denote such a foliation as $I\ni t\mapsto \Sigma_t$. Every leaf is a
$d$-dimensional embedded hypersurface, and we further assume that all leaves $\Sigma_t$
arise from embeddings of a single $d$-dimensional manifold $\Sigma$. It follows readily
from the implicit function theorem and the non-vanishing of the differential $dT$ that
each $p\in M$ has a chart neighborhood $U$ such that in local coordinates $\Sigma_t\cap U$
(if non-empty) consists of the points $(t,y^1,\ldots,y^d)$ in the chart range. Such adapted
coordinates are not unique. If $y$ and $y'$ 
are two such coordinate systems defined on an open set $U \subset M$, then 
both are related by a diffeomorphism of the form 
${y'}^0 = \chi^0(y^0), {y'}^a = \chi^a(y)= \chi^a(t,x)$, $a=1,\ldots,d$.  
By the implicit function theorem we also view $x^a(y)$ to be locally 
known and such that $\tilde y^{\alpha} = y^{\alpha}(t(\tilde y), x(\tilde y))$, 
for all $\tilde y^{\alpha}$. Here and below we also write $y^{\alpha}, 
\alpha =0,1,\ldots,d$, for $y = (y^0,y^a)$.

Two foliations ${I} \ni t \mapsto \Sigma_t$, and 
${I'} \ni t' \mapsto \Sigma'_{t'}$, defined on $M$ are called {\it equivalent} if there is a diffeomorphism 
sending the leaves of one into the leaves of the other. 
For simplicity we consider only smooth, orientation preserving
diffeomorphisms $\chi : M \ra M$ in the component 
of the identity,  that reduce to the identity outside a compact set. They form a group with respect to composition.
Sequences of 
diffeomorphisms and the concomitant topological 
considerations will not enter. 
For short, we just write ${\rm Diff}(M)$ for the resulting 
group of diffeomorphisms. 

In local charts, 
we identify points with their coordinates, and write alternatively 
$\chi(y)$ and $y'$ for the image point of $y \in U$. The 
differential $d\chi_y$   maps the tangent space at $y$ into the 
one at $y'$ and is written as $\dd {y'}^{\alpha}/\dd y^{\gamma}$. 
Similarly, for the inverse $\chi^{-1}: U' \ra U$, the image 
of $y' \in U'$ is written alternatively as $\chi^{-1}(y')$ and $y$. 
For the differentials one has $d(\chi\inv)_{y'}  = [d \chi_y]^{-1}$. 
In the $1\!+\!d$ decomposition we write $\chi^0$, $\chi^a$ and 
$(\chi^{-1})^0$, $(\chi^{-1})^a$ for the projections 
of $\chi$  and $\chi^{-1}$ onto an adapted coordinate 
basis, and whenever unambiguous we abbreviate those as $t'$, ${x'}^a$ 
and $t$, $x^a$, respectively. In this notation a generic 
$\chi \in {\rm Diff}(M)$ changes both the leaves of the foliation 
and the coordinatization of the hypersurfaces: 
\begin{equation} 
\label{fdiffeo1} 
t \mapsto \Sigma_t  \;\; 
\mbox{is mapped into} \;\;t' \mapsto \Sigma'_{t'}
\quad \mbox{by} \;\;  
t' = \chi^0(t,x),\, {x'}^a= \chi^a(t,x)\,.
\end{equation}
 By the above definition two such foliations are 
 equivalent. 
However, the adapted coordinates of one are not adapted to the other. 
This is to be contrasted with the subgroup ${\rm Diff}(\{\Sigma\}) 
\subset {\rm Diff}(M)$ of {\it foliation preserving diffeomorphisms}    
\begin{equation}
\label{fdiffeo2} 
\chi \in {\rm Diff}(\{\Sigma\}) \;\;\;\mbox{iff} \;\;\;
t' = \chi^0(t)\,,\;\;{x'}^a = \chi^a(t,x) \,.
\end{equation}
 As noted 
before, this is the maximal subgroup that maps adapted 
coordinates of a given foliation into each other; merely the 
labeling of the leaves and their {coordinatization} changes.  
The Jacobian matrix in the $1+d$ decomposition is then upper 
triangular. We reserve the notation ${\rm Diff}(\Sigma)$
for the subgroup of $t${-}independent diffeomorphisms ${x'}^a = 
\chi^a(x)$ of $\Sigma$. 
\medskip

{\bf Block decomposition of $1\!+\!d$ differentials.} 
The diffeomorphisms in $1\!+\!d$ form of course still form a group
under concatenation. Concatenating 
$(t', {x'}^a) = (\chi^0(t,x), \chi^a(t,x))$ 
with $(t'', {x''}^a) = ({\chi'}^0(t',x'), {\chi'}^a(t',x'))$ 
gives $(t'', {x''}^a) = ( (\chi' \circ \chi)^0(t,x)$, 
$(\chi' \circ \chi)^a(t,x))$, where
$(\chi' \circ \chi)^0(t,x) = {\chi'}^0( \chi^0(t,x), \chi^a(t,x))$
and $(\chi' \circ \chi)^a(t,x) = {\chi'}^a( \chi^0(t,x), \chi^a(t,x))$.
The defining relations for the inverse $\chi^{-1}$ of $\chi$ therefore are 
$(\chi^{-1})^0( \chi^0(t,x), \chi^b(t,x)) = t$,
$(\chi^{-1})^a( \chi^0(t,x), \chi^b(t,x)) = x^a$. 
In general, the temporal or spatial component of $\chi^{-1}$ 
also depends on the spatial or temporal component of $\chi$. 
An exception are diffeomorphisms trivial in one component, 
$(t,x^a) \mapsto (\chi^0(t,x), x^a)$ or 
$(t,x^a) \mapsto (t, \chi^a(t,x)))$, where the inverses 
depend only parametrically on $x^a$ or $t$, respectively.

Next, consider the composition of the differentials. 
Written in $1\!+\!d$ block form one has 
\be 
\label{diff1}
\frac{\dd y^{\gamma}}{\dd {y'}^{\alpha}} = 
\left( \begin{array}{cc} 
\dfrac{\dd t}{\dd t'} & 
\dfrac{\dd x^c}{\dd t'} 
\\[3mm] 
\dfrac{\dd t}{\dd {x'}^a} & 
\dfrac{\dd x^c}{\dd {x'}^a} 
\end{array} \right) \,,
\sspace 
\frac{\dd {y'}^{\alpha}}{\dd y^{\gamma}} = 
\left( \begin{array}{cc} 
\dfrac{\dd t'}{\dd t} & 
\dfrac{\dd {x'}^a}{\dd t} 
\\[3mm] 
\dfrac{\dd t'}{\dd x^c} & 
\dfrac{\dd {x'}^a}{\dd x^c} 
\end{array} \right) \,.
\ee
The chain rule 
$(\dd {y'}^{\gamma}/\dd y^{\beta})(\dd {y''}^{\alpha}/\dd {y'}^{\gamma})
= (\dd {y''}^{\alpha}/\dd y^{\beta})$ decomposes into blocks 
according to 
\ba 
\label{chain}
\frac{\dd t''}{\dd t} \is 
\frac{\dd t''}{\dd t'} \frac{\dd t'}{\dd t} 
+ \frac{\dd t''}{\dd {x'}^c} \frac{\dd {x'}^c}{\dd t} \,,
\nonum
\frac{\dd {x''}^a}{\dd t} \is 
\frac{\dd {x''}^a}{\dd t'} \frac{\dd t'}{\dd t} 
+ \frac{\dd {x''}^a}{\dd {x'}^c} \frac{\dd {x'}^c}{\dd t} \,,
\nonum
\frac{\dd t''}{\dd x^b} \is 
\frac{\dd t''}{\dd t'} \frac{\dd t'}{\dd x^b} 
+ \frac{\dd t''}{\dd {x'}^c} \frac{\dd {x'}^c}{\dd x^b} \,,
\nonum
\frac{\dd {x''}^a}{\dd x^b} \is 
\frac{\dd {x''}^a}{\dd t'} \frac{\dd t'}{\dd x^b} 
+ \frac{\dd {x''}^a}{\dd {x'}^c} \frac{\dd {x'}^c}{\dd x^b} \,.
\ea
As a consequence, the familiar inversion formula for 
the full Jacobian matrices (\ref{diff1}) does not project to the blocks. 
Systematically one would want to express the components of 
$\dd y^{\alpha}/\dd {y'}^{\beta}$ in terms of the components 
of $\dd {y'}^{\alpha}/\dd y^{\beta}$. To do so we specialize 
(\ref{chain}) to coinciding initial and final variables 
and swap the role of the primed and the unprimed fields. 
Combining the resulting equations pairwise gives 
\ba 
\label{inv1}
\frac{\dd x^c}{\dd {x'}^b} 
\bigg( \frac{\dd {x'}^a}{\dd x^c} 
- \Big(\frac{\dd t'}{\dd t} \Big)^{-1} 
\frac{\dd {x'}^a}{\dd t} \frac{\dd t'}{\dd x^c} \bigg) 
\is \delta_b^a\,,
\nonum
\frac{\dd x^c}{\dd t'} 
\bigg( \frac{\dd {x'}^a}{\dd x^c} 
- \Big(\frac{\dd t'}{\dd t} \Big)^{-1} \,
\frac{\dd {x'}^a}{\dd t} \frac{\dd t'}{\dd x^c} \bigg)
\is - \Big(\frac{\dd t'}{\dd t} \Big)^{-1} \,
\frac{\dd {x'}^a}{\dd t}\,. 
\ea 
The inverse of the matrix in brackets can be expressed 
in terms of the matrix inverse of $\dd {x'}^a/\dd x^b$ 
via the formula for rank one perturbations (Sherman-Morrison). 
Writing $Y_b^a$ for the result the desired inversion formulas 
read 
\ba
\label{inv2}
\frac{\dd x^a}{\dd {x'}^b} \is Y_b^a\,,
\nonum
\frac{\dd x^a}{\dd t'} \is - \Big( \frac{\dd t'}{\dd t} \Big)^{-1} \,
\frac{\dd {x'}^b}{\dd t} \, Y^a_b\,,
\nonum
\frac{\dd t}{\dd t'} \is \Big( \frac{\dd t'}{\dd t} \Big)^{-1}
+ \Big( \frac{\dd t'}{\dd t} \Big)^{-2}\,
\frac{\dd {x'}^d}{\dd t}\,Y^c_d \,\frac{\dd t'}{\dd x^c}\,,
\nonum
\frac{\dd t}{\dd {x'}^a} \is - \Big( \frac{\dd t'}{\dd t} \Big)^{-1}
Y^c_a \, \frac{\dd t'}{\dd x^c}\,. 
\ea
In general all components mix under inversion. Upper or lower 
block diagonal Jacobian matrices remain so, as required.
Only for direct product diffeomorphism $t'= \chi^0(t), 
\,{x'}^a = \chi^a(x)$ does (\ref{inv2}) reduce to the 
simple variants $\dd x^a/\dd {x'}^b = [(\dd {x'}/\dd x)^{-1}]_b^a$,
$\dd t/\dd t' = (\dd t'/\dd t)^{-1}$, directly entailed
by the implicit function theorem.  

In summary, the differentials $d\chi_y $ and $d(\chi^{-1})_{y'}  
= [d\chi_y ]^{-1}$ of a generic diffeomorphisms $\chi \in {\rm Diff}(M)$, 
admit a block decomposition whose composition and inverse 
is governed by the relations (\ref{chain}) and (\ref{inv2}).  
The advantage of this crude decomposition is that no 
metric structure is required.
\medskip

%%%%%%%%%%%%%%%%%%%%%%%%%%%%%%%%%%%%%%%%%%%%%%%%%%%%%%%%%%%%%%%%%  

{\bf Metric geometry of the foliations.}
We now consider the manifold $M$ to be equipped with a
pseudo-Riemannian metric $g^{\eps_g}$, which we take to
be smooth and similar to $(\eps_g, +, \ldots, +)$, $\eps_g =\mp 1$. For Lorentzian signature global hyperbolicity of $(M,g^-)$ is the instrumental 
condition. This entails that $M$ may be foliated by Cauchy slices, the existence of smooth temporal functions (see below), and 
the attainability of the $N^a =0$ gauge \cite{Shiftzerogauge}. 
Systematic expositions of the Lorentzian $1\!+\!d$ projection
formalism in metric geometry can be found in many textbooks,
see e.g.~\cite{Gbook}.
A  {\itshape temporal function}  in this context is a smooth function
$T: M \ra \R$ with a timelike gradient $d T$, interpreted
as a one-form $d T = (\dd T/\dd y^{\alpha}) dy^{\alpha}$. The associated vector field $g_-^{\alpha\beta} \dd_{\beta} T$  (with $g_-^{\alpha\beta}$ the
components of the inverse of $g^-_{\alpha\beta}$) is past pointing. 
Importantly, any globally hyperbolic spacetime admits a temporal 
function such that any level surface $\Sigma_t = \{ y \in M\,|\, 
T(y)~=~t~\}$ is a Cauchy surface \cite{Shiftzerogauge}. 
All level surfaces are diffeomorphic to a fixed manifold $\Sigma$, and $M$ itself is diffeomorphic to $\bbR\times \Sigma$.
This sets the relevant notion of foliation and we assume that all equivalent foliations are of this form. As a consequence the relevant foliation changing diffeomorphisms are of the form \eqref{tripletransf} for $\epsilon_g=-1$. 
For Riemannian metrics on manifolds $M$ diffeomorphic to
$\bbR\times \Sigma$ we shall continue to use the term `temporal
function' for a smooth function $T$ with a nowhere vanishing gradient. The equivalence of foliations will be defined through diffeomorphisms of type \eqref{tripletransf} for $\epsilon_g=+1$. 
The existence of a temporal function amounts to `time' orientability, and we assume  $\Sigma$ to be orientable as well (consistent with the assumed orientability of $M$). 
To fix the notation and to highlight the dependence on the signature parameter
$\eps_g \in \{\mp 1\}$, we  display the main relations of the
(Arnowitt-Deser-Misner) ADM formalism. For readability's sake we omit
the $\eps_g$ sub- or superscript in $g_{\alpha\beta}^{\eps_g}$ or
$g_{\eps_g}^{\alpha\beta}$ in the following.

For a fixed temporal function $T$ and the associated foliation 
$I \ni t \mapsto \Sigma_t$, one may identify $T$ with $t$
and write $\dd_{\alpha} t$ for the components of $d T$.   
In terms of them we set 
\begin{equation} 
\label{1d1}
g^{\alpha\beta} \dd_{\alpha} t \dd_{\beta} t =: \eps_g N^{-2}\,,
\quad 
m^{\alpha} := \eps_g N^2 g^{\alpha\beta} \dd_{\beta} t\,.
\end{equation}
The first equation defines the lapse $N$, the second defines a 
vector conjugate to the temporal gradient, $m^{\alpha} \dd_{\alpha} t
=1$. Note that $N$ is scalar and $m^{\alpha}$ a vector as 
long as $T$ is held fixed. Further $m^{\alpha} \dd_{\alpha}$ 
has unit coefficient along $\dd_t$ and 
\begin{equation}
m^{\alpha} \dd_{\alpha} = \dd_t - N^a \dd_a\,,
\end{equation}
defines the shift $N^a$. In terms of $m^{\alpha}, \dd_{\alpha} t$ 
projectors tangential and transversal to the leaves 
of the foliation are defined by 
\begin{equation}
\label{1d2} 
\Sigma_{\alpha}^{\;\;\beta} := \delta_{\alpha}^{\beta} - 
\dd_{\alpha} t \,m^{\beta}\,, \quad 
T_{\alpha}^{\;\;\beta} := \dd_{\alpha}t \,m^{\beta}\,.
\end{equation}
We write $\wg_{\alpha\beta} := \Sigma_{\alpha}^{\; \delta} 
\Sigma_{\beta}^{\; \gamma} g_{\delta\gamma}$ for the induced
metric on $\Sigma_t$. Since $m^{\alpha} \Sigma_{\alpha}^{\; \beta} =0$,
the natural derivative transversal to the leaves of the foliation 
is $e_0:= \cL_m = \dd_t - \cL_{\vec{N}}$, where $\cL_{\vec{N}}$ is the 
$d$-dimensional Lie derivative in the direction of $N^a$. 
When acting on scalars we write $e_0 = e_0^{\alpha} \dd_{\alpha}$,
so that $e_0^{\alpha} = m^{\alpha}$. The tangential derivatives acting 
on scalars are
\begin{equation}
\label{1d3}  
e_{\alpha}^a \frac{\dd}{\dd x^a} = \Sigma_{\alpha}^{\; \beta} 
\dd_{\beta} = \dd_{\alpha} - \dd_{\alpha} t \,e_0\,,
\quad 
\frac{\dd}{ \dd x^a} = e_a^{\alpha} \dd_{\alpha}\,.
\end{equation}
which defines the coefficient matrices $e_a^{\alpha}$ 
and $e_{\alpha}^a$. They are such that 
\ba
\label{1d4} 
&& e_a^{\alpha} \, e^b_{\alpha} = \delta_a^b\,,  
\sspace \;\;
\Sigma_{\alpha}^{\;\;\beta} = e^a_{\alpha} \,e_a^{\beta} \,,
\nonum
&& e_a^{\alpha} := g^{\alpha\beta} \wg_{ab} \,e^b_{\beta}\,,\quad 
g_{\alpha\beta} \,e^{\alpha}_a \,m^{\beta} =0 = 
g^{\alpha\beta} e_{\alpha}^a \,\dd_{\beta}t \,,
\ea
which express the orthogonality and completeness of the component
fields. By (\ref{1d1}), (\ref{1d2}), (\ref{1d4}) the metric and its 
inverse take the block diagonal form  
\ba 
\label{1d5}
g_{\alpha\beta} \is \eps_g N^2 \dd_{\alpha} t \dd_{\beta} t 
+ \wg_{ab} e^a_{\alpha} e^b_{\beta}\,,
\nonum
g^{\alpha\beta} \is \eps_g N^{-2} m^{\alpha} m^{\beta} 
+ \wg^{ab} e_a^{\alpha} e_b^{\beta}\,,
\ea
where $\wg^{ac} \wg_{cb} = \delta^a_b$. Further $\det g = \eps_g N^2 
\det \wg$. For a fixed temporal function in addition to 
$N,N^a$ also $\wg_{ab}$ is a scalar.

The description in terms of the embedding relations
$\tilde y^{\alpha} = y^{\alpha}(t(\tilde y), x(\tilde y))$ 
is now secondary, but still carries over
\begin{subequations}
\label{1d7}
\begin{eqnarray}
\label{1d7a}
&& \dd_{\alpha} t = \frac{\dd t}{\dd y^{\alpha}}\,,\sspace 
m^{\alpha} = \frac{\dd y^{\alpha}}{\dd t} - N^a e_a^{\alpha}\,,
\\
\label{1d7b}
&& e_a^{\alpha} = \frac{\dd y^{\alpha}}{\dd x^a} \,, 
\sspace e_{\alpha}^a = \frac{\dd x^a}{\dd y^{\alpha}} + N^a \dd_{\alpha} t 
\,,	
\end{eqnarray}
\end{subequations}
where $t(y)$ is the given temporal function and $x^a(y)$ is defined by the 
implicit function theorem. The left pair of relations holds by     
 definition. Further $\dd y^{\alpha} /\dd t - m^{\alpha}$ is 
orthogonal to $\dd_{\alpha} t$ and thus tangent to $\Sigma_t$. As such 
it can be written in the form $N^a e_a^{\alpha}$, which 
gives the second relation in \eqref{1d7a}. The orthogonality 
(\ref{1d4}) then provides the second relation in \eqref{1d7b}. 
The 1-forms $e^a = dx^a + N^a dt$ span the cotangent 
space of $\Sigma$, while $\dd_a = e_a^{\alpha} \dd_{\alpha}$ span the
tangent space. The full coordinate 1-forms and associated 
differentials are given by 
\ba
\label{1d8}
dy^{\alpha} \is m^{\alpha} dt + e^{\alpha}_a e^a\,,\,\sspace 
\;e^a = dx^a + N^a dt\,,
\nonum 
\frac{\dd}{\dd y^{\alpha}} \is \dd_{\alpha} t \,e_0 + e_{\alpha}^a \dd_a\,, 
\sspace e_0 = \dd_0 - N^a \dd_a\,.
\ea
The one forms $(N dt, e^a)$ and dual vector fields $(N^{-1} e_0, \dd_a)$
form a moving frame which we refer to as the {\it foliation frame}.
As long as the coordinate functions $t: U \ra \R$ and $x^a: U \ra \R^d$
are kept fixed the description is independent of the choice of embedding
coordinates $y^{\alpha}$.

%%%%%%%%%%%%%%%%%%%%%%%%%%%%%%%%%%%%%%%%%%%%%%%%%%%%%%%%%%%%%%%%%%%%
\section{Lapse-Wick rotation for metrics with preferred foliation}  

The framework developed here is primarily intended for {\it generic}
foliated metric geometries, where other notions of Wick rotation
do not apply. For spacetimes with isometries and/or a preferred foliation
the complexification induced by the lapse-Wick rotation often
differs in subtle and instructive ways from other notions of Wick
rotation and we outline the differences in this appendix for a few
examples. 

{\bf Minkowski space and static spacetimes.} The standard Wick
rotation for fields on Minkowski space is part of the pertinent
architecture of relativistic quantum field theories and can
be extended to static spacetimes \cite{Jaffe1,Jaffe2}. For simplicity
we restrict our comments here to Minkowski space. Most text book
treatments take Lorentzian signature and the Feynman propagator
as basic. The Wick rotation is then introduced as a deformation of
the integration contour in the time component of the momentum integration.  
In itself this is clearly limited to perturbation theory and even the
consequences for higher order diagrams are rarely discussed.  
A Wick rotation in time $t \mapsto e^{- i \th} t$ is used in
\cite{DuncanQFT}, p.328, and leads to a damping integrand in a 
(scalar selfinteracting) functional integral. The ensued
free Green's function does, however, for small $\th >0$ {\it not} induce
the Feynman $i \eps$ prescription but rather a variant originally
introduced by Zimmermann \cite{Zimmermann}. It is this notion of
Wick rotation that arises
by specialization of the lapse-Wick rotation to Minkowski space. 

To see how this comes about 
we use the line element $\eta_{\mu\nu} dy^{\mu} dy^{\nu} = - N_0^2 dt^2
+ \delta_{ab} dx^a dx^b$, for a constant lapse-like parameter $N_0$.
This, of course, is also an example of a spacetime with a preferred
foliation, where $N^a \equiv 0$. The hyperbolic slicing \cite{Gbook}  
is an another relevant foliation and would lead to a different notion
of lapse-Wick rotation. Either of them can however be studied in
generic non-fiducial foliations along the lines described in Section~2.
Using the standard foliation, the $N_0 \mapsto e^{- i \th} N_0$
lapse-Wick-rotated free Hessian (\ref{chess2}) with $V = m^2$ reads
\be
\label{Mink1} 
- i \Delta_{\th} = - e^{ i \th} (N_0^{-1} \dd_t)^2 +
e^{ - i \th} \delta^{ab} \dd_a \dd_b - e^{ - i \th} m^2\,.
\ee
The defining relation for the Green's function $-\Delta_{\th} G_{\th} =
\1$, is readily solved by Fourier transform and results in 
\be 
\label{Mink2}
G_{\th}(p_0, p) = \frac{ i e^{ - i \th}}{p_0^2 - e^{ -2 i \th} (p^2 + m^2)}\,,
\quad \th \in (0,\pi)\,,
\ee
where $p = (p_1, \ldots, p_d)$ is the spatial momentum vector,
$p^2 = \delta^{ab} p_a p_b$, and we set $N_0=1$ after the rotation.
For $\th = \pi/2$ this gives the Euclidean
propagator $1/(p_E^2 + m^2)$, $p_E^2 = p_0^2 + p^2$.  For $\th \ra 0^+$ the
behavior is $G_{\th}(p_0, p) = i(1+ O(\th))/[ p_0^2- p^2 - m^2
  + 2 i \th( p^2 + m^2)]$. With $2 \th = \eps$ this is precisely
the defining relation Eq.~(1.1) for Zimmermann's propagator
\cite{Zimmermann}. Compared to Feynman's
prescription this is `as if' $\eps \mapsto \eps\,( p^2 + m^2)$ has
been made dependent on the spatial momentum-squared. The expression
(\ref{Mink2}) extends Zimmermann's (distributional Lorentz
signature) propagator into the Euclidean regime. The qualitative
properties however remain the same for all $\th \in (0,\pi)$.
In particular, one has the crucial bounds 
\ba
\label{Mink3}
\frac{1}{p_E^2 + m^2} \leq | G_{\th}(p_0, p) |
\leq \frac{1}{\sin \th} \frac{1}{ p_E^2 + m^2}\,.
\ea 
As shown in \cite{Zimmermann}, (see also \cite{DuncanQFT}, p.618)
this has the important consequence
of rendering all (with Feynman's $i\eps$ prescription) conditionally
convergent integrals {\it absolutely} convergent. Hence the Euclidean
power counting theorems can be applied to the diagrams evaluated with
the Green's function (\ref{Mink2}). On the other hand, the distributional
$\th \ra 0^+$ limit is well defined and (re-)produces the desired
Lorentz invariant results, to all orders of renormalized perturbation theory
\cite{Zimmermann}. 
\medskip

{\bf De Sitter and Friedmann-Lema\^{i}tre spacetimes.} Another
important class of spacetimes
with a preferred foliation are Friedmann-Lema\^{i}tre cosmologies with
line element $g^{\rm FL}_{\mu\nu} dy^{\mu} d^{\nu} y=
- N(t)^2 dt^2  + a(t)^2 \delta_{ab} dx^a dx^b$, in $1\!+\!d$ dimensions.
Here we focus on the
spatially flat case for simplicity and keep the lapse $N(t)$ so as to
maintain temporal reparameterization invariance.  Cosmological time
corresponds to fixing $N(t) =1$, conformal time to the choice
$N(t) = a(t)$, etc.. Here one can see the main problem encountered with
a Wick rotation in time: it depends on the choice of time variable
and in general will render the scale factor a (possibly multi-valued)
complex function of it, which may or may not have
the desired two real sections. 
A detailed discussion of Wick rotations in time in the context of
cosmological path integrals can be found in \cite{FRWWick1, FRWWick2}.  
The Wick rotation in the lapse $N(t) \mapsto e^{-i \th} N(t)$ does
not depend on the choice of time and leads to the a complexified
metric on a {\it real} manifold, $(g^{\th}_{\mu\nu})^{\rm FL} dy^{\mu}
dy^{\nu} = - e^{ - 2 i \th} N(t)^2 dt^2 + a(t)^2 \delta_{ab} dx^a dx^b$.
When used in the scalar field action (\ref{Saction}) the general
results of Section 3 apply. In particular, both the linearized
and the nonlinear admissibility of the complexified metric continue
to hold also in non-fiducial foliations, even if the latter look
`unnatural' compared to the default foliation with $N^a \equiv 0$. 
That is, when using (\ref{swick1}) in a functional integral the
damping of the integrand is a foliation independent property, even if
the lapse-Wick-rotation itself is not.

This feature rests on the
positivity of the Euclidean signature action, it does not carry over
to situations where the Euclidean action is not bounded from below.
For example, in the Friedmann-Lema\^{i}tre mini-superspace action
(reduction of the Einstein-Hilbert action minimally
coupled to a self-interacting scalar field) the pattern
(\ref{swick1}) still applies, $S_{\th}^{\rm mini} = \cos \th
\,S_-^{\rm mini} + i \sin \th \,S_+^{\rm mini}$, but since
$S_+^{\rm mini}$ is not bounded from below, the notion of
admissibility from \cite{LS,KSWick} is not directly applicable.
The unboundedness of $S_+^{\rm mini}$ of course reflects the conformal factor
instability of the Euclidean Einstein-Hilbert action, and it needs
to be addressed independently, for example by including quadratic
curvature scalars. The relation $S_{\th} = \cos \th \,S_-
+ i \sin \,\th S_+$ itself applies to the $1+d$ form of the
Einstein-Hilbert action (Gibbons-Hawking action) as well. 

De Sitter space admits a flat slicing which is natural in
cosmological applications (but covers only part of the manifold).
Formally $a(t) = e^{t H}$ in the above Friedmann-Lema\^{i}tre line
element (but de Sitter space does not have a curvature singularity).
The lapse-Wick rotation then produces
$(g^{\th})^{\rm dS}_{\mu\nu} dy^{\mu} d^{\nu} y=
- e^{- 2 i \th}N_0^2 dt^2  + e^{2 Ht} \delta_{ab} dx^a dx^b$. 
It connects part of de Sitter space ($\th=0$) to the upper sheet of the
two-sheeted hyperbolid ($\th = \pi/2$).
If one were to emulate the transition by a Wick flip in
time $t \mapsto \pm i t$ also
the Hubble constant would need to be complexified $H \mapsto \mp i H$.  
Note that the Riemannian space obtained  is different from the
round sphere one finds by a Wick flip in time starting from
static coordinates or the global closed slicing,
see e.g.~\cite{dScomplex}.   
To illustrate the viability of the     
lapse-Wick rotation in this context we present (without derivation)
the lapse-Wick rotated heat kernel. This is the fundamental solution
of the heat-type equation $(\dd_s - \Delta_{\th})K_s(t,x;t',x') =0$,
$s >0$, where $\Delta_{\th}$ is obtained from (\ref{chess2}) by
specialization to the above flat slicing de Sitter metric and acts on
the first pair of arguments in $K_s$. The result is for
$\th \in (0,\pi)$ 
\ba
\label{dS1} 
&\!\!& K_s^{\th}(t,x;t',x') = (-i e^{i\th})^d H^{d+1}
\int_0^{\infty} \! d \om \, c(\om) \,e^{ s i e^{i\th} H^2[d^2/4 + \om^2]}
\,\Omega_{\om}\big( d_{\th}(t,x;t',x') \big)\,,
\nonum
&\!\!& c(\om) = \frac{1}{(2\pi)^{d+1}} \bigg|
\frac{\Gamma( i \om + d/2)}{ \Gamma(i \om) } \bigg|^2\,,
\quad \Omega_{\om}(\xi) =
(2\pi)^{ (d+1)/2} \frac{\cP^{-(d-1)/2}_{-1/2 + i \om}(\xi)}%
{ (\xi^1 -1)^{(d-1)/4}}\,.
\ea      
Here $\cP_{\mu}^{\nu}(z)$ is an associated Legendre function,
which has a branch cut from $-\infty$ to $1$. Further $c(\om)$
is the Harish-Chandra $c$-function for ${\rm SO}_0(1,d+1)/{\rm SO}(d+1)$.  
Finally, $d_{\th}$ is the embedding distance given by
\be
\label{dS2}
d_{\th}(t,x;t',x') = \cosh H(t-t') -
\frac{H^2}{2} e^{2 i \th} e^{(t+t') H} |x - x'|^2\,.
\ee
For $\th = \pi/2$ this coincides with the known heat kernel on
the upper sheet of the two-sheeted hyperboloid. A (near) de
Sitter counterpart is desirable and is often schematically
used (see e.g.~\cite{dScomplex}); the above expression provides
a mathematically valid construction.  The point to
stress is that the coordinates $(t,x)$, $(t',x')$ stay real,
the Wick rotation occurs through the phase $e^{i \th}$. Moreover
(\ref{dS1}) is manifestly well-defined for all $\th \in (0,\pi)$.
In particular, the value of $d_{\th}$ stays away from the branch cut,
and the exponent $s i e^{i\th} H^2[d^2/4 + \om^2]$ of the spectral value
has a negative real part for all $s, \om >0$ and $\th \in (0,\pi)$.  
The lapse-Wick rotated Green's function can be obtained from (\ref{dS2})
via a phase modified Laplace transform. 

The above result is a special case of a lapse-Wick rotated heat kernel
that can be defined on a generic foliated metric manifold without
isometries. It remains well-defined into the near Lorentzian regime
($\th>0$ small). The strict Lorentzian limit $\th \ra 0^+$ can be
taken under well defined traces \cite{Wickheatk}.

% ----------------------------------------------------------------------
%		Bibliography
% ----------------------------------------------------------------------

\end{document}